\documentclass[3p,amsmath,amssymb,compress]{elsarticle}

\usepackage[colorlinks]{hyperref}
\usepackage{amsmath}
\usepackage{graphicx}
\usepackage{latexsym}
\usepackage{amsmath,amssymb}
\usepackage{bm} 
\usepackage{xcolor}
\usepackage{stackrel}
\usepackage{accents}
\usepackage{braket}

\journal{Annals of Physics: contribution to the Konstantin B. Efetov Memorial Special Issue}

\numberwithin{equation}{section}

\begin{document}
\begin{frontmatter}	
\title{Exact solution of the topological symplectic Kondo problem}

\author[MPI]{Elio~J. K\"onig}
\author[BNL]{Alexei M. Tsvelik}

\address[MPI]{Max Planck Institute for Solid State Research, D-70569 Stuttgart, Germany}
\address[BNL]{Condensed Matter Physics and Materials Science Division, Brookhaven National Laboratory, Upton, New York 11973, USA}

\begin{abstract}
The Kondo effect is an archetypical phenomenon in the physics of strongly correlated electron systems. Recent attention has focused on the application of Kondo physics to quantum information science by exploiting overscreened Kondo impurities with residual anyon-like impurity entropy. While this physics was proposed in the fine-tuned multi-channel Kondo setup or in the Majorana-based topological Kondo effect, we here study the Kondo effect with symplectic symmetry Sp(2k) and present details about the implementation 
which importantly only involves conventional s-wave superconductivity coupled to an array of resonant levels and neither requires perfect channel symmetry nor Majorana fermions. We carefully discuss the role of perturbations and show that a global Zeeman drives the system to a 2-channel SU(k) fixed point.
Exact results for the residual entropy, specific heat and magnetization are derived using the thermodynamic Bethe Ansatz for Sp(2k). This solution not only proves the existence of a quantum critical ground state with anyon-like Hilbert space dimension, but also a particularly weak non-Fermi liquid behavior at criticality. We interpret the weakness of non-analyticities as a manifestation of suppressed density of states at the impurity causing only a very weak connection of putative anyons and conduction electrons. Given this weak connection, the simplicity of the design and the stability of the effect, we conjecture that the symplectic Kondo effect may be particularly suitable for quantum information applications.
\end{abstract}

\begin{keyword}
Kondo effect, Bethe Ansatz, Anyons, Quantum criticality, Quantum Dots
\end{keyword}

\end{frontmatter}

\tableofcontents

\section{Introduction}

The design of quantum devices and harvesting their properties for quantum technological utilization is a prime application of present day condensed matter research. In this context, topological quantum computation \cite{NayakDasSarma,PachosBook2012} is particularly appealing, because it allows for intrinsically robust storage and processing of information by braiding of anyons. The latter are low-energy quasiparticles which display generic exchange statistics, an irrational Hilbert space dimension (``quantum dimension''), and occur in strongly interacting many-body systems. 
The Kondo effect\cite{Hewson} is the archetypal problem in the research on strongly correlated quantum matter and therefore, maybe unsurprisingly, topological quantum computation\cite{BondersonNayak2008} has very recently been proposed on the basis of this platform\cite{LopesSela2020,Komijani2020,GabaySela2022}, too.

Historically, the Kondo effect first appeared in the context of magnetic impurities \cite{Sarachik1964} submerged in the Fermi sea of conduction electrons in metals and was marked by a characteristic minimum of the temperature dependence of the resistivity. 
It is well known that once the zero energy density of states of the bulk electrons can be taken as energy independent, the single impurity  model   effectively maps to a one-dimensional problem of chiral fermions of the form\cite{Andrei1980,Vigman1980,TsvelickWiegmann1983Review,AndreiLowenstein1983}:
\begin{eqnarray}
\mathcal H = i v_F\int dx \sum_\sigma c_\sigma^\dagger\partial_x c_\sigma + g \sum_{A \in \mathfrak g} J^A (0) \hat S^A, \label{1Dmodel}
\end{eqnarray}
where $x$ denotes the radial direction, $c_x$ are fermionic annihilation operators, $J^A(x) = \sum_\sigma c_{\sigma}^\dagger(x) \tau^A_{\sigma \sigma'} c_{ \sigma'}(x)$  
and $\hat S^A$ are components of the impurity spin {(we will henceforth set the Fermi velocity $v_F = 1$).} 
{As a matter of fact, the chiral Hamiltonian can be directly realized} in topological electronic systems with gapped bulk  where the boundary electronic modes are chiral. This fact is especially important for quantum information science applications since the corresponding Kondo model remains solvable even for the case of many impurities. 

In the original problem of magnetic impurities, the generators $\tau^A$ are taken from the $\mathfrak g = \mathfrak{su}(2)$ Lie algebra and the impurity spin is in the $\hat S^A$ fundamental representation of $\mathfrak g$. However the generalization to higher spin representations as well as to $\mathfrak g = \mathfrak{su}(n)$, where $ \sigma,\sigma' = 1,\dots. n$ with $n >2$, is not only theoretically appealing \cite{Coleman1983}, but also experimentally relevant describing both magnetic impurities of Ce and Yb
\cite{TsvelickWiegmann1983Review, Schlottmann1989}
and mesoscopic quantum devices~\cite{KellerGoldhaberGordon2014}. In the latter case, the role of the impurity atom is taken by a quantum dot in the charging energy dominated regime. The mesoscopic platform also allows to realize multichannel Kondo physics \cite{PotokGoldhaberGordon2007,KellerGoldhaberGordon2015,IftikharPierre2015, IftikharPierre2018,PouseGoldhaberGordon2021}, i.e.~Eq.~\eqref{1Dmodel} in which fermionic fields $c_{\sigma,a}(x)$ carry an additional orbital index $a = 1, \dots, k$ with $k>1$. While for $k = 1$ and fundamental representation of the impurity spin the ground state of the $\mathfrak su(n)$ Kondo problem is a Fermi liquid with a perfect screening of the impurity, the overscreened SU(2) multichannel problem with $k >2S$ is frustrated as the model is indeterminate which of the $k$ channels should screen the spin-S impurity~\cite{Nozieres1980}.  If 
the coupling to different channels is the same, the system is quantum critical, with a characteristic residual impurity entropy. The ground state degeneracy 
\begin{equation}
    g_k = \frac{\sin \left(\frac{(2S+1) \pi }{k+2}\right)}{\sin \left(\frac{\pi }{k+2}\right)} \label{eq:GroundstateSequence}
\end{equation}
responsible for this zero temperature entropy\cite{AndreiDestri1984,TsvelickWiegmann1985,Tsvelick1985,AffleckLudwig1993} corresponds for spin $S=1/2$ to the quantum dimensions of anyons of Ising~\cite{EmeryKivelson1992} ($k = 2$), Fibonacci ($k = 3$), and $\mathbb Z_3$ parafermion~\cite{KarkiMora2022} ($k = 4$) type.

\begin{figure}
    \centering
    \includegraphics{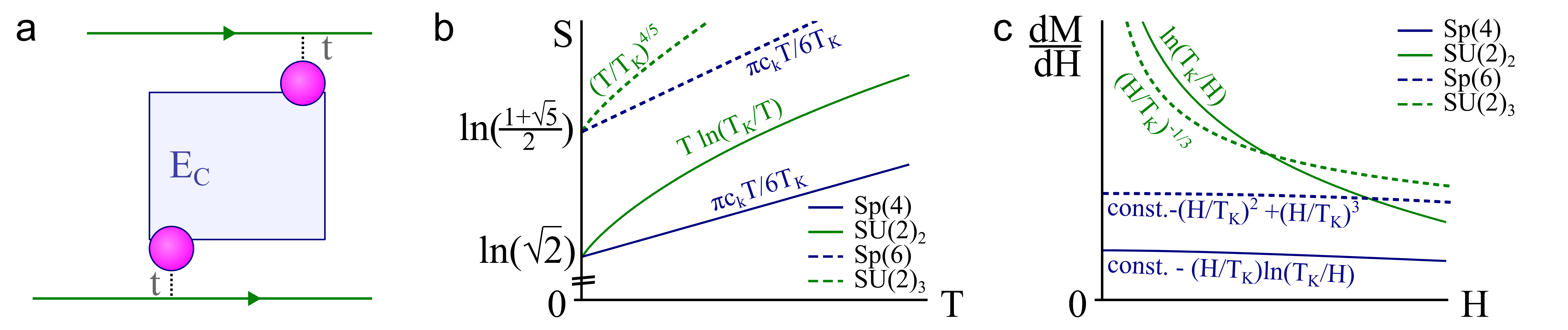}
    \caption{a) A sketch of the proposed Kondo device for $k=2$. The blue rectangle is a superconducting island with charging energy $E_C$. The pink dots are quantum dots with discrete single-particle spectrum of energy levels proximitized to the island. Electrons from the dots can tunnel to the chiral edges (in green) containing chiral electrons. b) Low-temperature entropy for Sp(2k) and k-channel SU(2) Kondo effect (both with impurity spin in the fundamental representation) for $k = 2$, $k=3$. Here {$c_k = c/2\cos[\pi/(2k + 2)]$ and $c$ is the central charge of the Sp$_1$(2k) theory and $T_K$ is the Kondo temperature.} c) The zero temperature susceptibility $\chi = dM/dH$  for the same models. For the symplectic Kondo effect, the leading behavior is Fermi liquid like.}
    \label{fig:SetupSummary}
\end{figure}

Another platform harboring anyonic quasiparticles is the ``topological Kondo effect'' proposed by B\'eri and Cooper\cite{BeriCooper2012,AltlandTsvelik2014,AltlandTsvelik2014b} and involves a topological Majorana Cooper pair box\cite{Fu2010, OregvonOppen2020} coupled to non-interacting metallic leads. In contrast to the multi-channel SU(2) case, this setup is stable to anisotropic coupling to different leads. Indeed, under renormalization group flow, it approaches Eq.~\eqref{1Dmodel} with $\mathfrak{g} = \mathfrak{o}(n)$ in the infrared  and thus realizes the orthogonal Kondo model which was theoretically studied since the 1990s as a simple formulation of two-channel Kondo physics~\cite{ColemanTsvelik1995}. One more situation where anyonic quasiparticles at a Kondo impurity are stable against imperfections occurs for strongly interacting leads~\cite{FieteNayak2008,KoenigTsvelik2020}. 

Here we study the ``symplectic topological Kondo effect'', i.e. Eq.~\eqref{1Dmodel} with $\mathfrak{g} = \mathfrak{sp}(2k)$, for which a mesoscopic implementation was recently proposed, see Ref.~\cite{LiVayrynen2022} and Sec.~\ref{sec:MesoSetup}. A common feature of this setup and of the topological Kondo setup is that neither require perfect channel symmetry which instead is dynamically generated in the infrared. At the same time, the symplectic Kondo setup does not assume the presence of Majorana fermions -- this distinguishing trait should facilitate a more imminent experimental realization as opposed to the topological Kondo effect. From the perspective of quantum materials, the study of a strongly correlated electron system with Sp(2k) symmetry is motivated by the relevance of SO(5)$ \sim $Sp(4) symmetry for the physics of high-$T_c$ superconductivity\cite{DemlerZhang2004} as well as the role of the symplectic symmetry of spin \cite{FlintColeman2008} in context of heavy-fermion superconductivity.

In this paper, we present the Bethe ansatz equations for the Sp(2k) symmetric Kondo impurity model and derive the integral equations for the impurity thermodynamics. These methods have been previously  applied to obtain exact results in SU(n) Kondo models\cite{Vigman1980,Andrei1980,TsvelickWiegmann1983Review,AndreiLowenstein1983,JerezZarand1998} (including different spin representations and number of channels\cite{AndreiDestri1984,TsvelickWiegmann1985,Tsvelick1985}). It was also applied to the Kondo problem of orthogonal symmetry\cite{AltlandTsvelik2014b,Buccheri2015}, but to the best of our knowledge, never to the symplectic Kondo problem. Consistently with results from conformal field theory\cite{Kimura2021}, the strong coupling analysis of the problem\cite{LiVayrynen2022} and Abelian bosoniziation \cite{MitchellAffleck2021,LibermanSela2021} [the latter being valid only for the special case $SO(5) \sim Sp(4)$], we find that the symplectic Kondo effect is characterized by a non-trivial ground state degeneracy $g_k$ which is the same as Eq.~\eqref{eq:GroundstateSequence} in the case $S = (k-1)/2$ or $S= 1/2$, as well as Fermi liquid behavior in the leading temperature dependence of specific heat and magnetic field dependence of magnetization. Additionally, we determine the leading non-Fermi liquid corrections occurring at higher order. 
The nearly Fermi-liquid like behavior is in sharp contrast to both multi channel SU(2) Kondo effect and the O(N) Kondo effect and indicates a smaller density of states near the impurity indicative of weaker coupling of anyons to the conduction bath. We believe that this feature should be favorable in the context of quantum information science. We also derive the local magnetization in the presence of a global Zeeman field and demonstrate 2-channel SU(k) Kondo behavior.

This paper is structured as follows: {Sec.~\ref{sec:The model } describes the model set up, the impact of perturbations and, to make the paper self-contained, a review about the mathematical structure of the symplectic group. } Sec.~\ref{sec:MainResults} contains the main results obtained using the Bethe Ansatz. Sec.~\ref{sec:BetheAnsatz} contains the derivation of the main results of the paper and may be skipped by readers interested only in the main physical effect. We conclude with an outlook section. Technical details are contained in the appendix.

\section{Model of symplectic Kondo effect and experimental tuning parameters}
\label{sec:The model }

In this section we summarize the experimental setup, partly reviewing results presented in Ref.~\cite{LiVayrynen2022}.

\subsection{Mesoscopic setup}
\label{sec:MesoSetup}

The symplectic impurity spin is implemented by means of the
following model for a quantum dot, Fig.~\ref{fig:SetupSummary} a):

\begin{align}
    \mathcal H_{\rm dot} & = E_C (2 \hat{N}_C + \hat n_d - N_g)^2 - \frac{\Delta}{2} \sum_{i = 1}^k [e^{-i \phi} d^\dagger_{i,\sigma} [\sigma_y]_{\sigma \sigma'} d^\dagger_{i, \sigma'} + H.c.]. \label{eq:Hdot}
\end{align}

Here, Einstein convention for spin summation is used,   $\hat n_d = \sum_{i} d^\dagger_{i, \sigma} d_{i \sigma}$ is the total charge in the edge states and $\hat N_C = -i \partial_\phi$ counts the number of Cooper pairs. The label $i = 1, \dots, k$ refers to the number of resonant zero energy states coupled to the superconducting island. 

Importantly, the dot Hamiltonian preserves the total charge $2 \hat{N}_C + \hat n_d = \hat N_{\rm tot}$, as well as the fermionic parity and spin on each of the $k$ resonant levels. This is physically realized when Cooper pairs are interconverted into two electrons only locally at each of the $k$ resonant levels which is satisfied when the distance {between the 
quantum wells {$i = 1, \dots, k$}}
exceeds the superconducting coherence length.

We here focus on the particle-hole invariant point $N_g = 1$. In the limit $0< E_C- \Delta  \ll E_C,\Delta$, the ground state of the quantum dot is in the odd parity sector, $2k$-fold degenerate, and labelled by the index and spin of the only singly occupied level, $\ket{i,\sigma}$. There are two smallest excited states at energy $E_C - \Delta$ above the ground state with one higher or lower unit of charge, respectively.

The $2k$ degenerate ground states are characterized by the occupation of a single fermion with spin $\sigma$ on one of the $k$ resonant levels.
For $t\ll E_C - \Delta$, the coupling to the leads 
\begin{equation}
    \mathcal H_{t} = \sum_{i =1}^k \sqrt{a} t c_{i, \sigma}(x = 0) d_{i,\sigma} + H.c.,
\end{equation}
($a$ is an ultraviolet length scale) is taken into account perturbatively by means of two virtual  processes corresponding to transitions  into one of the two states with one additional and one missing charge and generates the following superexchange interaction: 
\begin{equation}
    \mathcal H_{\rm superexchange} = \frac{g}{2} \ket{a}\bra{b} c^\dagger_{a'} c_{b'} [\delta_{ab'} \delta_{a'b} - (\sigma_y)_{aa'} (\sigma_y)_{b'b}].
\end{equation}
Here, $a = (i,\sigma)$ is a multi-index and $g = 4t^2/(E_C - \Delta)$.
Under application  of the Fierz identity, the addition of these two processes leads to Eq.~(\ref{1Dmodel}) with $\mathfrak{g} =\mathfrak{sp}(2k)$. 

\subsection{Symplectic Lie algebra and root system}
\label{sec:Roots}

\begin{figure}[t]
    \centering
    \includegraphics[width = \textwidth]{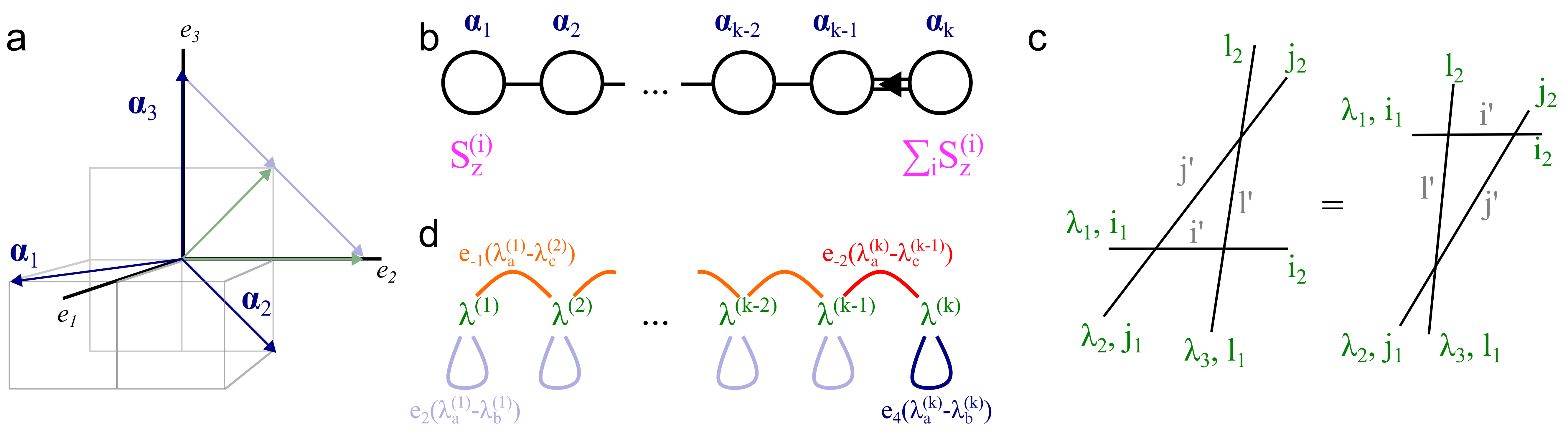}
    \caption{a) Root system of Sp(2k) at $k = 3$, corresponding to Eq.~(\ref{eq:RootSystem}). We also present all the positive roots in the $\hat e_k - \hat e_{k-1}$ plane (includes the green arrows). b) Dynkin diagram for the symplectic group Sp(2k). Note the double line connecting the last two roots (the diagram for $k = 2$ contains only these last two roots). The first and last two roots are associated to the local and total spin polarization. c) Yang-Baxter equation. d) Graphical illustration of the Bethe equations \eqref{eq:BA1},\eqref{eq:BA2}. }
    \label{fig:Dynkin}
\end{figure}

It is helpful for the remainder of the paper to review some basic properties of the theory of Lie algebras. Readers who are familiar with this mathematics or only interested in the physical results may want to skip this section.

We remind the reader that elements of the symplectic Lie algebra have the property
\begin{equation}
    {\tau^A}^T = - \sigma_y \tau^A \sigma_y, \label{eq:SympLie}
\end{equation}
and analogously for the impurity operators $\hat S^A$ for $A \in \mathfrak{sp}(2k)$. 

The Cartan subgroup is the maximal Abelian subgroup of the Lie group. In the present case of $2k \times 2k$ matrices sufficing Eq.~\eqref{eq:SympLie} it is physically reasonable to choose $H_i = \hat \Pi_i \otimes \sigma_z$ as Cartan generators ($\hat \Pi_i$ projects on the $i$th resonant level). Of course, in the case of $k = 1$ corresponding to Sp(2) $\simeq$ SU(2), there is only one Cartan generator corresponding to the spin quantum number.

The root system of a Lie group physically corresponds to the generalization of magnetic raising and lowering operators from the case of one Cartan generator [as in SU(2)] to the more generic case. Specifically, we consider the following set of eigenvectors of the Cartan generators $E_{j}^+ = \hat \Pi_j \sigma_+$ ($\sigma_\pm = \sigma_x \pm i \sigma_y)$, $E_{lm}^+ = \sigma_z [\hat e_l \hat e_m^T+ l\leftrightarrow m] + [\hat e_l \hat e_m^T- l\leftrightarrow m]$ ($[\hat e_i] = \delta_{ij}$, $i, j = 1, \dots k$ and $l \neq m$)  with the properties 
\begin{subequations}
\begin{align}
    [H_i, E_{j}^+] &= \underbrace{2 \delta_{ij}}_{\alpha^{(j)}_i} E_{j}^+,\\
    [H_i, E_{lm}^+] &= \underbrace{[\delta_{li}  - \delta_{mi}]}_{\alpha^{(lm)}_i} E^+_{lm}.
\end{align}
\label{eq:EigenvaluesLie}
\end{subequations}
The vectors of eigenvalues,  $\boldsymbol{\alpha}^{(j)} = (\alpha^{(j)}_1, \dots, \alpha^{(j)}_k),\boldsymbol{\alpha}^{(lm)}  = (\alpha^{(lm)}_1, \dots,  \alpha^{(lm)}_k)$ are called roots. Clearly, the operators $E_{j}^+ = \Pi_j \sigma_+, E_{lm}^+ = \sigma_z [\hat e_l \hat e_m^T+ l\leftrightarrow m] + [\hat e_l \hat e_m^T- l\leftrightarrow m]$ play the role of magnetic raising operators, while their complex conjugates are lowering operators with opposite roots.

Positive roots can be defined by the criterion that their first non-zero entry is positive, i.e. keeping only $\boldsymbol{\alpha}^{(lm)} $ with $l<m$. The $k$-dimensional base of the system of positive roots is given by 

\begin{align}
    \lbrace \boldsymbol{\alpha}_1,\boldsymbol{\alpha}_2 \dots , \boldsymbol{\alpha}_k\rbrace  = \lbrace \hat e_1 - \hat e_2, \hat e_2 - \hat e_3, \dots, \hat e_{k-1} - \hat e_k,  2 \hat e_k \rbrace \hat = \lbrace E_{12}^+, E_{23}^+, \dots, E_{k-1,k}^+,  E_k^+ \rbrace, \label{eq:RootSystem}
\end{align}
where the collection of raising operators on the right of $\hat =$ is in one-to-one correspondence with the roots on the left.

Clearly, the angles between adjacent roots is $120^\circ$ for the first $k-2$ pairs of roots and $135^\circ$ for the last pair, and $90^\circ$ otherwise, see Fig.~\ref{fig:Dynkin} a). This is encoded by a single line, a double line and no line in the Dynkin diagram, Fig.~\ref{fig:Dynkin} b).

\subsection{Perturbations}
In the low-energy limit, the unperturbed Kondo Hamiltonian Eq.~\eqref{1Dmodel} is supplemented by ``Zeeman''-like terms
\begin{equation}
    \delta \mathcal H = \sum_{A \in \mathfrak{sp}(2k)} H^A \hat S^A + \sum_{A \in \mathfrak{su}(2k) \backslash \mathfrak{sp}(2k)} \tilde H^A \hat S^A.
\end{equation}
Here we summarize the microscopic origin and physical meaning of various relevant such perturbations.

First, we concentrate on perturbations which generate Zeeman-like field $H^A$ within the symplectic group. These include local physical Zeeman fields on each given island, as well as antisymmetric hybridizations on the island (no sum over $i,j$ in this section),
\begin{subequations}
\begin{align}
    \delta \mathcal H_{\rm Zeeman} &= -H_i d^\dagger_i \sigma_z d_i \longrightarrow - H_i \hat \Pi_i \otimes \sigma_z, \\
    \delta \mathcal  H_{\rm \pi/2-hopping} &= -i td^\dagger_i d_j +h.c. \stackrel{i = 1, j = 2}{\longrightarrow} t \left (\begin{array}{cccc}
            0 & -i & \dots &0  \\
            i & 0 & \dots &0 \\
            \vdots & \vdots & \ddots & \vdots \\
            0 & 0 & \dots & 0
        \end{array}\right) \otimes \mathbf 1_\sigma.
\end{align}
\end{subequations}
The expressions to the right of the arrow indicate the effective operators in the impurity Hilbert space, which in the fundamental representation is given by the direct product of $k \times k$ matrices and spin.
Note that the $\pi/2-$phase of the hopping element in the second line can not be merely absorbed into a redefinition of $d_j$, as it would alter the pairing term in Eq.~\eqref{eq:Hdot}.

Second, perturbations which generate Zeeman-like fields $\tilde H^A$ outside the symplectic group include  symmetric hybridizations on the island, crossed Andreev reflections and local variations of the order parameter field
\begin{subequations}
\begin{align}
        \delta\mathcal  H_{\rm hopping} &= - t_0 \sum_\sigma d^\dagger_{i, \sigma} d_{j, \sigma} + H.c. \stackrel{i = 1, j = 2}{\longrightarrow} - t_0 \left (\begin{array}{cccc}
            0 & 1 & \dots &0  \\
            1 & 0 & \dots &0 \\
            \vdots & \vdots & \ddots & \vdots \\
            0 & 0 & \dots & 0
        \end{array}\right) \otimes \mathbf 1_\sigma, \\
        \delta\mathcal  H_{\rm CAR} &= \Delta_{\rm CAR} e^{- i \phi} d^\dagger_{i} \sigma_y d_{j}^\dagger + H.c. \stackrel{i = 1, j = 2}{\longrightarrow} (\Delta_{\rm CAR} + \Delta^*_{\rm CAR}) \left (\begin{array}{cccc}
            0 & 1 & \dots &0  \\
            1 & 0 & \dots &0 \\
            \vdots & \vdots & \ddots & \vdots \\
            0 & 0 & \dots & 0
        \end{array}\right) \otimes \mathbf 1_\sigma, \\
        \delta \mathcal H_{\rm \Delta} &=2\delta \Delta e^{- i \phi} d^\dagger_{i} \sigma_y d_{j}^\dagger + H.c. \stackrel{i = 1 =j }{\longrightarrow} (\delta \Delta + \delta \Delta^*) \left (\begin{array}{cccc}
            0 & 0 & \dots &0  \\
            0 & 1 & \dots &0 \\
            \vdots & \vdots & \ddots & \vdots \\
            0 & 0 & \dots & 1
        \end{array}\right) \otimes \mathbf 1_\sigma.
\end{align}
\end{subequations}

Finally, one may wonder about the impact of a local chemical potential $\delta \mathcal H_{\mu_j} = - \mu_j  d^\dagger_{j} d_j $.
As the ground states $\ket{i,\sigma}$ of Eq.~\eqref{eq:Hdot} are given by a BCS state for all levels except $i$  (i.e. it is a coherent superposition of zero and two particles), the expectation value of the local number operator $d^\dagger_{j} d_j$ is unity for all $l = 1, \dots, k$, independently on $j$ or $i$. As such, $\delta \mathcal H_{\mu_j} \rightarrow - \mu_i \mathbf 1_k \otimes \mathbf 1_\sigma$ is a featureless addition to the Hamiltonian. This demonstrates the stability of the symplectic Kondo device with respect to small detuning of the levels on the quantum dots.

In the remainder of the paper, we will focus on the physically most relevant and tunable perturbation, i.e. Zeeman fields $H_i$. 
Given the definition of roots presented in the previous section \ref{sec:Roots}, it is easy to see that that applying a local magnetic field to one resonant level (e.g. to level $i = 1$) leads to a non-trivial commutator with $E_{12}^+$ and commutes with all other eigenvectors of Eq.~\eqref{eq:RootSystem}, while a homogeneous field $H_i = H, \forall i$ commutes with all eigenvectors except $E_k^+$. In this sense, a local Zeeman field
corresponds to the first (short) root of the Dynkin diagram, while the homogeneous field is associated to the last (long) root, see Fig.~\ref{fig:Dynkin} b).

\subsection{Limit of SU(k) 2-channel Kondo effect.}
\label{sec:SUklimit}

Here we discuss the effective model in the presence of a homogeneous Zeeman field $H \ll E_C,\Delta$ acting on the impurity only and for an impurity in the fundamental representation of Sp(2k). It is possible to use a $2k \times 2k$ matrix representation for $\hat S^A \in \lbrace i\hat  A \otimes \hat{\mathbf 1}_\sigma + \hat{\vec S} \otimes \hat{\vec \sigma} \rbrace$, where $A$ are antisymmetric $k\times k$ matrices and the components of $\vec S = (S_x,S_y,S_z)$ are symmetric  $k\times k$ matrices. The Kondo part of the Hamiltonian Eq.~\eqref{1Dmodel} becomes
\begin{equation}
    \mathcal H_K = g \sum_A \left \lbrace  [c^\dagger(0) i A^A \otimes \mathbf 1_\sigma c(0)] (i\hat  A^A \otimes \hat{\mathbf 1}_\sigma) + \sum_{\mu = x,y,z} [c^\dagger(0) S_\mu^A \otimes \sigma_\mu c(0)] (\hat  S_\mu^A \otimes\hat \sigma_\mu) \right\rbrace.
\end{equation}

In the limit of energies below $H$ we project $\mathcal H_K$ by means of $\hat \Pi_\uparrow = (\mathbf 1 + \hat \sigma_z)/2$ to obtain
\begin{align}
\hat \Pi_\uparrow \mathcal H_K \hat \Pi_\uparrow & = g \sum_A \left \lbrace  [c^\dagger(0) i A^A \otimes \mathbf 1_\sigma c(0)] i\hat  A^A + [c^\dagger(0)  S_z^A \otimes \sigma_z c(0)]\hat  S_z^A   \right\rbrace \otimes\hat \Pi_\uparrow \notag\\
&=g \sum_A \left \lbrace  [c^\dagger(0) \left (\begin{array}{cc}
    H^A & 0 \\
    0   & -(H^T)^A
\end{array} \right)_\sigma c(0)] \hat  H^A   \right\rbrace\otimes \hat \Pi_\uparrow .
\end{align}
The Hermitian $k\times k$ matrices (operators) $H^A = S_z^A + i A^A$ ($\hat H^A = \hat S_z^A + i \hat A^A$) are generators of U(k), and under performing a particle-hole transformation $c_\downarrow \leftrightarrow c_\downarrow^\dagger$ it is apparent that two identical electronic leads compete for screening the U(k) impurity. 

\section{Main results of the Bethe Ansatz calculation}
\label{sec:MainResults}

 Our main result is the full set of thermodynamic Bethe Ansatz (TBA) equations for the Sp(2k) Kondo model. These equations contain the full information of the thermodynamic properties. We used these equations to extract some concrete results for the thermodynamics, some of which have been known from other methods such as conformal field theory. We have formulated conditions for quantum criticality and have calculated the ground state entropy and non-Fermi liquid corrections to the specific heat and magnetic susceptibilities. The quantum critical ground state has a finite nontrivial entropy indicating the presence of anyon zero energy modes. For the full Sp(2k) symmetry the  non-analytic corrections to the specific heat and magnetization are subdominant.  However, the application of  magnetic field causes a crossover from Sp$_1$(2k) to SU$_2$(k) QCP where the non-Fermi liquid corrections are dominant.

Specifically, when the impurity spin is in the fundamental representation of Sp(2k) we find in the absence of external magnetic fields a free energy (see Sec.~\ref{sec:BetheAnsatz} for details)
\begin{equation}\label{eq:F}
    F \stackrel{T \ll T_K}{\simeq } - \frac{\pi c}{12}\frac{T^2}{E_F} + \frac{1}{N} \left [- T \ln(g_k) - \frac{\pi c_k}{12}  \frac{T^2}{T_K} - \dots - \text{const.} \times T \left(\frac{T}{T_K} \right)^{k+1} \ln(T_K/T)\right].
\end{equation}
{Here, $c_k = c/{2 \cos\left (\frac{\pi}{2k + 2} \right)}$, where $c$ is the central charge of the Sp(2k)$_1$ theory.}
The impurity free energy in the square brackets contains a zero temperature entropy consistent with an anyonic quantum dimension
\begin{equation}\label{eq:DegeneracySummarySec}
    g_k = 2\cos\Big(\frac{\pi}{k+2}\Big),
\end{equation}
i.e. to the ground state degeneracy as for the SU(2) $k-$channel Kondo effect for $S = 1/2$, Eq.~\eqref{eq:GroundstateSequence}. Moreover, we find that the leading finite temperature corrections lead to a Sommerfeld coefficient $C/T \simeq \pi/6 [c/E_F + c_k/NT_K]$, in which the role of the Fermi energy $E_F$ is taken by the Kondo temperature $c T_K/c_k$, otherwise it is given by the standard expression. 
In contrast to the SU(2) $k-$channel Kondo effect, the leading finite temperature corrections are Fermi liquid like, Fig.~\ref{fig:SetupSummary} b).

The zero temperature impurity magnetization in the presence of a small local Zeeman field $H_i \ll T_K$ takes the form 
\begin{equation}
    M_i = \alpha_1 H_i/T_K + \alpha_3 (H_i/T_K)^3+ \dots + \beta (H_i/T_K)^{k+1} \times \begin{cases} 1, & \quad k \text{ odd,} \\
     \ln(T_K/H_i), & \quad k \text{ even.} \end{cases} \label{eq:MagSummary}
\end{equation}
where $\alpha_{1,3,\dots}, \beta$ are dimensionless numbers determined implicitly in Sec.~\ref{sec:BetheAnsatz}. Note that logarithmic corrections are absent if $k$ is odd, however even powers in $H_i/T_K$ appear (neither are present in a Fermi liquid state). We also find
\begin{equation}
    M_{\rm tot} = \alpha_{\rm tot} H/T_K,
\end{equation}
without any non-analytical corrections in the case of  small global magnetic field $H_i = H, \forall i$. 

In the case $H_i \gg T_K$, the TBA restores the perturbative result $M_i \propto \ln (H_i/T_K)$, and similar for the global Zeeman field $H$. 

For this last case, $H_i = H, \forall i$, we also study the response to weak additional local Zeeman fields, i.e. $H_i = H + \delta H \delta_{i,1}$ for $T_K \gg H \gg \delta H$ and find
\begin{equation}
     M_1 = \begin{cases} \bar \alpha_1\left (\frac{\delta H}{T_K}\frac{H^2}{E_F^2} \right) \ln\left (\frac{T_K}{\delta H}\frac{E_F^2}{H^2} \right), & k = 2, \\
     \bar{\alpha}_1  \left (\frac{\delta H}{T_K}\frac{H^2}{E_F^2} \right) + \dots + \bar{\beta} \left (\frac{\delta H}{T_K} \right)^{k/2} \left (\frac{H^2}{E_F^2} \right)^{\frac{2+k}{2}}, & k > 2, \text{odd},\\
     \bar{\alpha}_1  \left (\frac{\delta H}{T_K}\frac{H^2}{E_F^2} \right) + \dots + \bar{\beta} \left (\frac{\delta H}{T_K} \right)^{k/2} \left (\frac{H^2}{E_F^2} \right)^{\frac{2+k}{2}} \ln\left (\frac{T_K}{\delta H}\frac{E_F^2}{H^2} \right), & k > 2, \text{even}.
     \end{cases} \label{eq:MagSummary2}
\end{equation}
(with numerical coefficients $\bar \alpha_1, \bar \beta$) which under the identification of $H_{\rm eff} = \delta H (H/E_F)^{\frac{2+k}{k}}$ is the same behavior of the $2-$channel SU(k) Kondo effect. 
This effect is indeed expected based on the mesoscopic implementation introduced above, under the assumption that all impurity sites are spin polarized, see Sec.~\ref{sec:SUklimit}. 
 
 Three comments about the magnetization $M_1$ are in order. The first one is that the Fermi liquid component has completely disappeared. The response to the "small" field $\delta H_1$ is singular. The same
singularity will appear in the Sommerfeld coefficient in the specific heat. The second comment is that the conditions for SU$_2$(k) QCP hold  even for $H \gg T_K$ and requires only a smallness of $H_{\rm eff}$. Third, we note that the two channel SU(k) Kondo problem~\cite{JerezZarand1998} has the same ground state degeneracy, Eq.~\eqref{eq:DegeneracySummarySec}, as the Sp(2k) Kondo effect while the leading thermodynamic corrections are Fermi liquid like for $k$ exceeding the number of channels. 

\section{Bethe Ansatz solution}
\label{sec:BetheAnsatz}

One can diagonalize  the one-dimensional Hamiltonian (\ref{1Dmodel}) by brute force, solving the many-body Schr\"odinger equation for particles on a ring of length $L =Na_0$ with periodic boundary conditions. This approach leads us to the intermediary problem of diagonalization of the factorized N-body scattering matrix:
\begin{subequations}
\begin{align}
 E &=\sum_j k_j, \label{eq:GeneralBetheEnergy}\\
 e^{ik_j L}e^{i\hat \phi(k_j)}\xi &= \hat S(k_j, k_1)...\hat S(k_j,k_a)...\hat S(k_j,k_N)\xi, \label{Smatrix}
\end{align}
\label{eq:BetheGeneral}
\end{subequations}
where $\xi$ is a suitably chosen component of the electron wave function,  
 $\hat \phi$ is the impurity scattering phase and $\hat S(k,k')$ is the 2-body scattering matrix as a function of momenta $k,k'$. The integrability is hidden in the fact that the $N$-body problem for arbitrary number of particles $N$ can be represented as a succession of 2-body scattering problems. This is possible provided the eigenvalues are invariant under permutations of the particle's momenta which is fulfilled when the two-body $S$-matrix satisfies the Yang-Baxter equation, Fig.~\ref{fig:Dynkin} c). 
 
\subsection{Bethe equations}

{We can skip the first step of diagonalizing the S-matrices, using the fact that} the general classification of Bethe ansatz (BA) equations (\ref{Smatrix}) for all simple Lie groups was given by Reshetikhin and Wiegmann \cite{ReshWieg87,PCF}. 
These results are presented in the form of a set of algebraic equations for the 
rapidities $\lambda_a^{(j)}$ (reparametrizations of momenta for the $a$-th particle of type $j$).  The general structure of the equations is determined by the group symmetry and reflects the Dynkin diagram for the given group. This, however, leaves a certain ambiguity, since models with the same symmetry may be realized for
different group representations. 

 Just as the general BA equations, the diagonalization of the Sp(2k) Kondo problem contains an equation for the energy
 
\begin{equation}
    E = -\frac{ N}{2a_0} \sum_{a=1}^{M_1}  p_2(\lambda_a^{(1)}) + \sum_{{j}=1}^k h_j M_j 
    \label{eq:Energy}
\end{equation}

where $p_n(\lambda) =  i \ln[e_n(\lambda)]$, $e_n(x) = ({x + in/2})/({x-i n/2})$ and the index $j = 1, \dots, k$ enumerates the simple roots of the symplectic group (with {``Zeeman fields''} $h_j$) and thereby different types of particles. Given that the Fermi velocity is set to unity, the average interparticle distance $a_0\sim L/N$ is inversely proportional to the Fermi energy.
In the $j$th sector, $M_j$ is the number of occupied states and physically encodes the magnetization corresponding to the $j$th root. We remind the reader that we identify $M_1$ ($h_1$) with local magnetization (local Zeeman field) of the mesoscopic setup, $M_k$ ($h_k$)  with the global magnetization (global Zeeman field) on the device. In addition to Eq.~\eqref{eq:Energy}, the total energy contains a contribution from the charge sector (defined as the part of the Hilbert space which is decoupled from the impurity), see Ref.~\cite{TsvelickWiegmann1983Review} for a review. 

 The Dynkin diagram for the symplectic group is displayed in Fig.~\ref{fig:Dynkin} b). The case $k = 2$ containing only two roots connected by a double line is somewhat different from the case $k \geq 3$, where the first $k-1$ roots are connected by single lines. This is reflected in the Bethe equations as follows: For $k = 2$, the generic equation~\eqref{Smatrix} for $S$-matrices becomes

\begin{subequations}
\begin{align}
     [e_2(\lambda_{a}^{(1)})]^N e_q(\lambda_{a}^{(1)} +1/g) &= \prod_{\substack{b = 1,\\ b\neq a}}^{M_1}e_2(\lambda_a^{(1)}-\lambda_b^{(1)})\prod_{c=1}^{M_2}e_{-2}(\lambda_a^{(1)}- \lambda_c^{(2)}),\\
1&= \prod_{\substack{b = 1,\\ b\neq a}}^{M_2}e_4(\lambda_a^{(2)}-\lambda_b^{(2)})\prod_{c=1}^{M_1}e_{-2}(\lambda_a^{(2)}- \lambda_c^{(1)}),
\end{align}
\label{eq:BA1}
\end{subequations}
while for $k \geq 3$
\begin{subequations}
\begin{align}
    [e_2(\lambda_{a}^{(1)})]^N e_q(\lambda_{a}^{(1)} +1/g) &= \prod_{\substack{b = 1,\\ b\neq a}}^{M_1}e_2(\lambda_a^{(1)}-\lambda_b^{(1)})\prod_{c=1}^{M_2}e_{-1}(\lambda_a^{(1)}- \lambda_c^{(2)}),\\
1&= \prod_{\substack{b = 1,\\ b\neq a}}^{M_{j}}e_2(\lambda_a^{(j)}-\lambda_b^{(j)})
\prod_{c=1}^{M_{{j-1}}}e_{-1}(\lambda_a^{(j)}- \lambda_{c}^{(j-1)}) 
\prod_{c=1}^{M_{j+1}} e_{-1}(\lambda_a^{(j)}- \lambda_c^{(j+1)}), \label{eq:BA2b}\\
1&= \prod_{\substack{b = 1,\\ b\neq a}}^{M_{k-1}}e_2(\lambda_a^{(k-1)}-\lambda_b^{(k-1)})
\prod_{c=1}^{M_{k-2}}e_{-1}(\lambda_a^{(k-1)}- \lambda_c^{(k-2)})
\prod_{c=1}^{M_k}
e_{-2}(\lambda_a^{(k-1)}- \lambda_c^{(k)}),\\
1&= \prod_{\substack{b = 1,\\ b\neq a}}^{M_k}e_4(\lambda_a^{(k)}-\lambda_b^{(k)})\prod_{c=1}^{M_{k-1}}e_{-2}(\lambda_a^{(k)}- \lambda_c^{(k-1)}),
\end{align}
\label{eq:BA2}
\end{subequations}
where Eq.~\eqref{eq:BA2b} is restricted to $j = 2, \dots, k-2$. Each $\lambda^{(j)}$ interacts with the set of rapidities $\lambda^{(j')}$ of the same root $j' = j$ and of adjacent roots $j' = j \pm 1$ as shown on the Dynkin diagram. 
The subscript of $e_n(\lambda^{(j)} - {\lambda^{(j\pm 1)}})$ encoding the scattering matrix is determined as follows: 
$n = -1$ if the roots are connected by a single line and $n = -2$ if they are connected by a double line. The subscript for the same root $j' = j$ is positive and is determined by its length: $n=2$ for short roots and $n=4$ for the long root, see Fig.~\ref{fig:Dynkin} d) 

 The  position of the driving term (the term on the left-hand side of Eqs.~(\ref{eq:BA1},\ref{eq:BA2}); the only term  containing information about {the Hamiltonian including the interaction with impurity}) associated with the bulk Hamiltonian is determined by the group representation of the current operator of the conduction electrons and is consistent with \cite{ReshWieg87,PCF}. 
 The values of $q$ are determined by the representation of the impurity spin {and the} position 
 among the equations {labelled by} ${j=1, \dots, k}$ is determined by  the occupation $n$  such that $n = 2{(j-1)} +q$. In (\ref{eq:BA2}) it stands in the first equation which corresponds to one occupied orbital; for ${j} <k$ occupied orbitals it would be placed in the ${j}$-th equation.  As we shall see, only  $q=1$ corresponds to quantum critical ground state {while $q = 2$ leads to a Fermi liquid state. We expect this behavior to hold}
independently of {the} position {of the driving term}.

\subsection{Thermodynamic Bethe Ansatz}

In this section, we derive the most important (i.e. defining) equation of the TBA for the symplectic Kondo problem, compactly written as Eq.~\eqref{eq:TBA} below.

In the thermodynamic limit $N \rightarrow \infty, M_j/N =$const the solutions of the discrete BA equations (\ref{eq:BA1},\ref{eq:BA2}) are grouped into strings~\cite{TsvelickWiegmann1983Review,VanTongeren2016}. 
In the given case we have two sets of string solutions. The complex rapidities $\lambda^{(j)}$ group in complexes of $n$ rapidities with common real part $\lambda^{(j,n)}$:
\begin{equation}
    \lambda^{(j)}_{\alpha,n,p} = \begin{cases} \lambda^{(j,n)}_{\alpha} + \frac{i}{2}(n+1 -2p) + \dots, \quad & \text{for } j = 1, \dots, k-1, \\ 
    \lambda^{(j,n)}_{\alpha} + i (n+1 -2p) + \dots, \quad & \text{for } j = k,
    \end{cases}
\end{equation}
with $n=1,2,...$ and $p = 1, \dots n$ and the ellipsis denote terms which are exponentially small in system size and vanish in the thermodynamic limit. In this notation, $\alpha = 1, \dots M_{nj}$ labels the occupied states for particles of type $(j,n)$ and $\sum_{n = 1}^\infty n M_{nj} = M_j$. 

Leaving details to ~\ref{sec:DerivationTBAEquations}, we derive the thermodynamic Bethe equations governing the density of particles (holes)  $\rho_{n}^{(j)}(\lambda)$ [$\tilde\rho_{n}^{(j)}(\lambda)$] of string centers $\lambda_\alpha^{(j,n)}$ ($j = 1, \dots k-1$) and analogously $\sigma_{n}^{(j)}(\lambda)$ [$\tilde\sigma_{n}^{(j)}(\lambda)$] for the string centers $\lambda_\alpha^{(k,n)}$. For the sake of a lighter notation it is further useful to define the functions $\rho_{n}^{(k)} = \frac{(1 + (-1)^{n})}{2} \sigma_{n/2}$ and analogously for holes. 

We take the product over imaginary parts of the strings to get the energy
\begin{equation}
E = -\frac{N}{2a_0} \sum_{j = 1}^k\sum_n \int d \lambda [p_{n+1}(\lambda) + (1-\delta_{n1})p_{n-1}(\lambda)] \rho_n^{(j)}(\lambda) + \sum_j \sum_n \int d\lambda h_j n \rho_n^{(j)}(\lambda), \label{eq:EnergyContinuum}
\end{equation}
and can use a compact notation for the $k$ equations (valid for any $k \geq 2$)
\begin{equation}
{ \delta_{j,1} \left [A_{n,1}(\lambda) + \frac{1}{N} \left \lbrace \begin{array}{c}
A_{n,1} * s(\lambda + 1/g) \\ 
A_{n,1} (\lambda + 1/g)
\end{array}  \right \rbrace \right ]  = \tilde \rho_n^{(j)} (\lambda) +  {\mathbf A}_{nm}^{jl}* \rho_{m}^{(l)}(\lambda), \label{eq:TBA} }
\end{equation}
where 
\begin{equation}
{\mathbf A}_{nm}^{jl} = A_{nm} * ( {C}_{jl} - \delta_{j,k} \delta_{l,k} d). \label{eq:Kernel}
\end{equation}
It is implied that the $j = k$ and $l =k$ components of ${\mathbf A}_{nm}^{jl}$ are evaluated for even $n$ only. In the curly brackets, the case $q = 1$ is the first line, the second line is $q = 2$. 
For the convolution of two functions we have
$
f*g(\lambda) = \int_{-\infty} ^{\infty}du f(\lambda -u)g(u),
$ 
and it is useful to define functions of $\lambda$ by its Fourier transform
$
f(\lambda) \equiv \frac{1}{2\pi}\int d\omega f(\omega)e^{-i\omega\lambda}
$:
\begin{subequations}
\begin{align}
A_{nm}(\omega) &= \coth(\vert \omega \vert/2) [e^{-\vert n- m \vert \vert \omega\vert/2} -e^{- (n+ m) \vert \omega\vert/2}], \label{eq:app:A} \\
C_{mn} (\omega) &= \delta_{mn}- s(\omega)[\delta_{m, n +1} + \delta_{m, n-1}],\\
s(\omega) &= \frac{1}{2\cosh(\omega/2)},\\
d(\omega) &= 1 - \frac{\coth(\omega)}{\coth(\omega/2)}. 
\label{eq:dDef}
\end{align}
\label{eq:DefsMainText}
\end{subequations}
While the matrix elements of $C_{jl}(\omega)$ are the same as those of $C_{mn}(\omega)$, the identity $C_{mn} (\omega) = [A^{-1}]_{mn}$ assumes summation over all positive integers $m,n$ (this identity is thus not valid for the summation over roots $j,l = 1, \dots k$). Note that for $n \geq 2,  A_{n,1} = A_{n,2} *s$, in this case we can write $A_{n,q}*s$ for both lines of the curly brackets in Eq.~\eqref{eq:TBA}. 

\subsection{Zero temperature solution}

The easiest way to ascertain the non-Abelian nature of the $q=1$ QCP is to solve the BA equations for T=0 in nonzero magnetic field. There are several such fields which opens a possibility of subtle manipulation of the device. By balancing the fields one can drive the system from the Sp(2k) QCP to 
fixed points with smaller symmetry including the trivial one corresponding to the Fermi liquid. One interesting possibility is to depopulate $\lambda^{(k)}$ rapidities which will drive the system to SU$_2$(k) QCP which has the same ground state degeneracy, but different set of irrelevant operators.

We here anticipate that only $\rho_2^{(j)}$ in  Eq.~\eqref{eq:TBA} are non-zero at zero temperature (this can be justified by analyzing the $T\rightarrow \infty$ limit of the finite temperature free energy and Bethe equations discussed in Sec.~\ref{sec:FiniteTemp} and \ref{sec:DerivationTBAEquations}). Application of magnetic fields create holes in these distributions. For general magnetic fields
\begin{subequations}
\begin{align}
\rho_2^{(j)}(\lambda) & = 0 \text{ for } \lambda< -B_j,\\
\tilde \rho_2^{(j)}(\lambda) & = 0 \text{ for } \lambda> -B_j,
\end{align}
\end{subequations}
where the limits $B_j$ are determined by the Zeeman fields $h_j$.
When $B_j \rightarrow \infty$, $\tilde \rho_2^{(j)}(\lambda) =0, \forall \lambda$, which is the ground state in the absence of Zeeman fields. We will present analytical results for limit cases when one of the fields is much larger than the others.

\subsubsection{A single non-zero Zeeman field}
We first consider the situation in which there is a single non-zero Zeeman field. Inversion of Eq.~\eqref{eq:TBA} (see ~\ref{app:ZeroT}) leads to
\begin{equation}
\rho_2^{(j)} + K_{jl}* \tilde \rho_2^{(l)} = [K_{j1} *A_{22}*s](\lambda) +  \frac{1}{N}[K_{j1} *A_{2q}*s](\lambda + 1/g),\label{eq:ZeroTInverseTBA}
\end{equation}
where 
\begin{equation}
K_{jl} =  \frac{e^{\vert \omega \vert}}{\sinh(\omega) \cosh([k+1] \omega/2)}\cosh([k+1- \max(j,l)] \omega/2)\sinh(\min(j,l) \omega/2). 
\end{equation}

In the limit when only one $B_j < \infty$, Eq.~\eqref{eq:ZeroTInverseTBA} becomes as single equation for $\rho_j, \tilde \rho_j$ which can be solved using the Wiener-Hopf method, in which $K_{jj}(\omega) = G_j^{(-)}(\omega)G_j^{(+)}(\omega)$ is split into two functions with poles only in lower and upper half-plane, respectively, see~\ref{app:ZeroT}. The relationship between $B_j$ and applied fields $h_j$ is determined using the bulk contribution to the magnetization.
In the scaling limit $E_F = 1/\pi a_0 \rightarrow \infty, g \rightarrow 0, T_K = E_F e^{- \frac{\pi}{(k +1)g }} \rightarrow \text{const.}$, the impurity magnetization $M_{\rm imp} \sim - \tilde \rho_2^{(j)} (\omega = 0) \vert_{\rm imp}$ is 

\begin{equation}
M_j\vert_{\rm imp} \sim -\frac{i}{2\pi  G_j^{(-)}(\omega =0)} \int d \omega' \frac{1}{ \omega' + i0}\frac{e^{i \omega' (1+k)\ln(T_K/h_j)/\pi} r_1^{(j)} (\omega')}{{G_j^{(+)}(\omega')} }, \label{eq:Mag}
\end{equation}
with 
\begin{equation}
    r_1^{(j)} (\omega) = \frac{\cosh[(k + 1 - j) \omega/2]}{\cosh[(k + 1) \omega/2]}\frac{\sinh(q \omega/2)}{\sinh(\omega)}.
\end{equation}
At $h_j \gg T_K$, the integral can be bent in the lower half-plane where $G^{(+)}$ has a branch cut. Hence we get an expansion in negative powers of the logarithm $\ln(h_j/T_K)$ corresponding to perturbation theory.

For the low-field behavior $h_j \ll T_K$ of the magnetization the integral contour can be bent into the upper half-plane. The leading low-field behavior is given by the poles of $r_1^{(j)}(\omega)$ which are closest to the real axis, i.e. $\omega = i {\pi}({k + 1})$. Non-analyticities stem from double poles in $r_1^{(j)}(\omega)$ and from poles which are incommensurate with the prefactor of the logarithm in the exponential function of Eq.~\eqref{eq:Mag}. In particular, the leading non-analytical corrections in Eq.~\eqref{eq:MagSummary} stem from $\omega = i \pi$. Note that these kind of poles and by consequence the non-Fermi liquid behavior are absent when $q = 2$ or $j = k$. 

\subsubsection{Two non-zero Zeeman fields, $h_j \gg h_l$. Solution for $j = k$.}

Having determined $\rho_2^{(j)}$, $\tilde \rho_2^{(j)}$, the leading behavior for $\rho_2^{(l)}$, $l \neq j$ can be straightforwardly determined from Eq.~\eqref{eq:TBA} which are effectively linear in the limit $\tilde \rho_2^{(l)} = 0$. In the case $j = k$, we now consider finite  $\tilde \rho_2^{(l)}$ corrections, this leads to SU(k) 2-channel Kondo behavior.

In the first approximation, again we use $\tilde \rho_2^{(l)} = 0$ and solve for $\tilde \rho_2^{(k)}$ following the strategy outlined in the previous section. As mentioned, this solution is Fermi-liquid like. As the second step, we write down the equation of $\rho_2^{(l)}$, $l < k$,
\begin{align}
\delta_{l,1} \left [A_{2,1}(\lambda) + \frac{1}{N} 
[A_{2,q} * s](\lambda + 1/g) \right ]  = \tilde \rho_2^{(l)} (\lambda) + \sum_{i = 1}^k [{A}_{22} * C^{li}* \rho_{2}^{(i)}](\lambda),
\end{align}
and invert the $(k-1)\times(k-1)$ matrix $C_{li}$ leading to
\begin{equation}
\rho_2^{(l)} + \sum_{i = 1}^{k-1}\bar K_{li}* \tilde \rho_2^{(i)} = [\bar K_{l1} *A_{22}*s](\lambda) +  \frac{1}{N}[\bar K_{l1} *A_{2q}*s](\lambda + 1/g) -\sum_{i = 1}^{k-1}[\bar K_{li} * A_{22} * C_{ik} * \rho_2^{(k)}](\lambda) ,\label{eq:ZeroTInverseTBA2}
\end{equation}
where 
\begin{equation}
\bar K_{li} =  \frac{e^{\vert \omega \vert}}{\sinh(\omega) \sinh(k\omega/2)} \sinh([k-  \max(i,l)]\omega/2) \sinh(\min(i,l) \omega/2). \label{eq:barK}
\end{equation}
Note that $\rho_2^{(k)}$ enters in a way similar to the bulk contribution $[\bar K_{l1} *A_{22}*s]$. By means of arguments outlined above (see~\ref{app:ZeroT} for details) it determines a relationship between $B_l$ and $h_l^{\rm eff} = h_l (h_k/E_F)^{(2+k)/k}$. Then, the solution of Eq.~\eqref{eq:ZeroTInverseTBA} in the scaling limit with $T_K \sim E_F e^{- 2\pi/(kg)}$ and using the Wiener-Hopf method with $\bar K_{ll} = D_l^{(-)}(\omega) D_l^{(+)}(\omega)$ leads to 
\begin{equation}
M_l\vert_{\rm imp} \sim -\frac{i}{2\pi  D_l^{(-)}(\omega =0)} \int d \omega' \frac{1}{ \omega' + i0}\frac{e^{i \omega' \frac{k}{2\pi}\ln(T_K/h_l^{\rm eff})} \bar r_1^{(l)} (\omega')}{{D_l^{(+)}(\omega')} }, \label{eq:Mag2}
\end{equation}
with 
\begin{equation}
    \bar r_1^{(l)} (\omega) = \frac{\sinh[(k-j) \omega/2]}{\sinh(k \omega /2)}\frac{\sinh(q \omega/2)}{\sinh(\omega)}.
\end{equation}

Again the leading small-field behavior is given by the pole closest to the real axis, which is $\omega = i 2 \pi/k$. For $k = 2$ and $q = 1$ this is a double pole leading to a the logarithmic dependence of magnetization on the effective field. In contrast, for $k>2$ the leading field dependence in Eq.~\eqref{eq:MagSummary2} is Fermi liquid like, and leading non-Fermi liquid corrections stem from $\omega = i \pi,$ which is present only for $q = 1$ and additionally it is a double pole for $k$ even. 

\subsection{Finite temperature solution, ground state degeneracy}
\label{sec:FiniteTemp}

In this section we will derive the equations which determine the BA thermodynamics. As it happens for integrable systems this description will come in the form of a system of nonlinear integral equations for the excitation energies at thermodynamic equilibrium (TBA equations).
It turns out that some features of the thermodynamics can be extracted from these equations without solving them. In this task we will be greatly helped by the fact that the bulk part of the solution describes the free electron gas (at least that part of it which is described by the Sp$_1$(2k) Wess-Zumino-Novikov-Witten model). 

\subsubsection{General result for the free energy}

In the universal scaling limit, the impurity free energy takes the form (details about these equations are relegated to~\ref{sec:DerivationTBAEquations}) 
\begin{align} 
F_{\rm imp} =& - T \sum_{j = 1}^{k-1} \int d\lambda \ln(1 + e^{\phi_1^{(j)}(\lambda)}) G_j^{(1)}[\lambda + (1+k)\ln(T/T_K)/\pi] \delta_{q,1} \notag \\
&- T \sum_{j = 1}^k \int d\lambda \ln(1 + e^{\phi_2^{(j)}(\lambda)}) F_j^{(q)}[\lambda + (1+k)\ln(T/T_K)/\pi] \label{eq:FimpFiniteT}
\end{align} 
with $T_K = \frac{2}{a_0}e^{- \frac{\pi}{(1+k) g}}$ and 
\begin{align}
G_j^{(1)}(\omega)
& = \frac{\sinh[(k-j)\omega/2]}{\sinh(k \omega/2)},\label{eq:Gj1}\\
F_j^{(1)}(\omega)
& = \frac{\cosh \left[{(k + 1-j) \omega }/{2}\right]}{2 \cosh \left({\omega }/{2}\right) \cosh \left[(k+1) \omega/2 \right]}. \label{eq:GFMaintext}
\end{align}
Note that $G_k^{(q)}(\omega)=0$, and that $F_j^{(1)}(\omega) = F_j^{(2)}(\omega) s(\omega)$. The functions $\phi_n^{(j)}$ are governed by the set of equations
 \begin{equation}
 - \sin\left ( \frac{\pi l}{2k+2}\right) e^{\frac{\pi  \lambda }{1+k}}\delta_{m,2} =  \phi_m^{(l)} + [\mathbf{C}_{mn}^{lj}
 -\delta_{lj} \delta_{mn}]*
 \ln (1 + e^{\phi_n^{(j)}}) \label{eq:PhiEq}
 \end{equation}
  where $\lim_{n \rightarrow \infty}\phi_n^{(j)}/n = h^{(j)}/T$, 
  where $\mathbf{C}_{mn}^{lj} \mathbf{A}_{nm'}^{jl'} = \delta_{mm'} \delta_{ll'}$ is the inverse of the kernel \eqref{eq:Kernel}, see~\ref{app:InverseKernel}.
   Physically, $\phi_n^{j}(\lambda) = \ln(\tilde \rho_n^{(j)}/\rho_n^{(j)}) \vert_{\lambda + (1 +k)\ln(a_0 T/2)/pi}$ is a measure of the relative occupation particles and holes.

{Equations~\eqref{eq:FimpFiniteT},\eqref{eq:GFMaintext},\eqref{eq:PhiEq}} contain all information about the impurity thermodynamics. In the present paper we will not undertake any numerical solution of these equations. Instead we restrict ourselves to the low temperature limit. The bulk free energy has the same form as Eq.~\eqref{eq:FimpFiniteT} for $q = 2$ and $T_K \rightarrow E_F$.

\subsubsection{Zero temperature entropy}

  All functions $G_j^{(q)}(x),F_j^{(q)}(x)$  have a peak at $x=0$ and fall off exponentially. When the temperature $T\ll T_K$ the peak shifts towards 
  $\lambda \rightarrow \infty$ where
  $\phi_2^{(j)}(x)$ go to $-\infty$. So the corresponding contributions to the impurity entropy vanish at zero temperature and we concentrate on the solutions $\phi_{1, \infty}^{(j)} = \phi_{1}^{(j)} (\lambda \rightarrow \infty), j = 1, \dots, k-1$  which are 
determined by (see~\ref{app:ZeroTDeg} for details)
\begin{equation}
0 = \phi_{1, \infty}^{(j)} + \frac{1}{2} \ln[(1 + e^{-\phi_{1, \infty}^{(j+1)}})(1 + e^{-\phi_{1, \infty}^{(j-1)}})]. \label{eq:ZeroTlambdas}
\end{equation}
These equations are similar, but not the same as the two channel problem~\cite{Tsvelick1985}. The solution is
\begin{equation}
(1 + e^{- \phi_{1,\infty}^{(j)}}) = \frac{\sin^2\left ( \frac{\pi (j + 1)}{k+2} \right)}{\sin^2\left ( \frac{\pi}{k+2} \right)},
\end{equation}
which is even under $j \rightarrow k-j$. Using this symmetry and employing the defining equations~\eqref{eq:ZeroTlambdas} we can simplify the lowest-temperature free energy
\begin{align}
F_{\rm imp}\vert_{T \rightarrow 0} &= - T \sum_{j = 1}^k  {\frac{k-j}{k}}\ln(1 + e^{\phi_{1, \infty}^{(j)}}) \notag \\
&= -\frac{T}{2} \Big [\frac{\ln(1 + e^{- \phi_{1,\infty}^{(1)}})}{2} + \frac{\ln(1 + e^{- \phi_{1,\infty}^{(k-1)}})}{2}\Big ] 
\end{align} 
 This readily leads to $F = - T \ln(g_k)$ with $g_k$ given in Eq.~\eqref{eq:DegeneracySummarySec}.

\subsubsection{Leading temperature corrections}

Now we can calculate the leading finite $T$ corrections to the entropy. We first discuss the leading temperature corrections which turn out to be Fermi liquid like. Importantly, the pole of $F_j^{(2)}$ is $\omega = \pm i \pi/(k+1)$ and, for $q = 2$, leads to the identical quadratic temperature dependence of bulk and impurity free energy. We identify the associated Sommerfeld coefficient with the central charge of the Sp$_1$(2k) theory. Also in the case $q = 1$ the leading pole of $F_j^{(1)}$ is $\omega = \pm i \pi/(k+1)$, yet its residue is altered by the additional factor $s(\omega)$. Thus, the Sommerfeld-coefficients of the impurity specific heat at different $q$ obey
\begin{equation}
    \frac{C/T \vert_{q = 1}}{C/T \vert_{q = 2}} = 2 \cos\left (\frac{\pi}{2k + 2}\right).
\end{equation}

 To discuss the non-analytical corrections we concentrate on $k = 2$.
From (\ref{eq:PhiEq}) 
we see that $\phi_1^{(1)}$ interpolates from its value 0 at $-\infty$ to its asymptote at $+\infty$, so that $\ln(1+e^{\phi_1^{(1)}})$ interpolates between $\ln 2$ and its saturation at around $\lambda =0$. From the same equation we see that the growth is exponential $\sim \exp(\pi\lambda)$ and this exponent cancels the exponential decay of the $G_j^{(1)}[\lambda + (3/\pi)\ln(T_K/T)]$ kernel in (\ref{eq:GFMaintext}) in the interval $(-(3/\pi)\ln(T_K/T), 0)$. So with logarithmic accuracy we get for the non-analytic contribution the expression 
\begin{eqnarray}
\delta F_{imp} \sim - (T^4/T_K^3)\ln(T_K/T).\label{nonanF}
\end{eqnarray}
{A similar argumentation holds for $k >2$ and leads to Eq.~\eqref{eq:F}}.

\section{Outlook}

In conclusion, we have derived exact results for the symplectic Kondo problem~\eqref{1Dmodel} using Bethe Ansatz technique. A synopsis of results is contained in Sec.~\ref{sec:MainResults}. We therefore refrain from a further summary and simply highlight that we found anyon-like ground state degeneracies combined with leading Fermi-liquid like behavior -- this is in sharp contrast to the overscreened Kondo effect of SU(2) or O(M) symmetry. Additionally, we have reviewed a mesoscopic implementation of the symplectic Kondo effect and carefully discussed the role of symmetry breaking fields.

We conclude with an outlook on future work. In the context of the quantum information theoretical interest~\cite{LopesSela2020, Komijani2020} of using emergent anyons at Kondo impurities for (measurement-only) topological quantum computation~\cite{BondersonNayak2008}, it has been proposed to study an array of quantum impurities along a channel of chiral conduction electrons~\cite{GabaySela2022,LotemGoldstein2022,LotemGoldstein2022b}. Since chiral fermions do not generate Ruderman–Kittel–Kasuya–Yosida interactions, we argue that effectively the impurities remain independent and thereby the problem remains integrable. A careful study of this setup will be left for a future publication.

\section{Acknowledgments}
It is a pleasure to thank Laura Classen, Yashar Komijani, Guangjie Li, Jukka V\"ayrynen for useful discussions. This work was supported by Office of Basic Energy Sciences, Material Sciences and Engineering Division, U.S. Department of Energy (DOE)
 under Contracts No. DE-SC0012704 (AMT). 
 
\appendix

\section{Derivation of the thermodynamic Bethe Ansatz equations}

\subsection{General strategy}

In this Appendix, we outline the procedure which leads from the discrete Bethe ansatz equations (\ref{eq:BA1},\ref{eq:BA2}) to their thermodynamic limit represented by the  equations for the densities (\ref{eq:TBA}).
We follow the 
standard procedure, see e.g.~\cite{TsvelickWiegmann1983Review,VanTongeren2016}, we will thus restrict ourselves by the demonstration of principle. 

 Namely, instead of the relatively complicated equations (\ref{eq:BA1},\ref{eq:BA2}) with many types of rapidities we will consider the simplest example of the 1D Bose gas with the delta function repulsion $V = c\delta(x)$ where all particles momenta are real. The discrete Bethe ansatz equations are 
 \begin{eqnarray}
 && e^{ik_j L}= \prod_{l\neq j}^N\frac{k_j - k_l +ic}{k_j -k_l -ic},\label{BG1}\\
 && E = \sum_{j=1}^Nk_j^2
 \end{eqnarray}
 Taking the logarithm of both parts of (\ref{BG1}) we obtain
 \begin{equation}
 k_jL = 2\pi n_j - 2\sum_{l=1}^N \tan^{-1}[(k_j-k_l)/c]. \label{BG2}
 \end{equation}
 Here the set of integer numbers $\{n_j\}$ characterizes the state of the system. In the thermodynamic limit the integer numbers are densely distributed; in a given eigenstate the ones which are "filled" are interspersed by ones which are "empty". The idea is to characterize the state by two continuum functions - the particle density $\rho(k)$ and hole density $\tilde\rho(k)$. These functions are not independent and related by the equation which we will now derive. As the first step we replace each sum over filled states  by an integral over the density of filled states:
 \begin{equation}
 \sum_l \tan^{-1}[(k_j - k_l)/c] \rightarrow L\int dp \rho(p) \tan^{-1}[(k_j -p)/c].
 \end{equation}
 Then we will augment the equation for physical momenta (\ref{BG2}) by the equation for the momenta of holes:
 \begin{equation}
 \tilde k_jL = 2\pi \tilde n_j - 2\sum_{l=1}^N \tan^{-1}[(\tilde k_j-k_l)/c], \label{BG3}
 \end{equation}
 where integer numbers $\tilde n_j$ fill the holes in the set $n_j$.  Combining  Eqs.(\ref{BG2},\ref{BG3}) together we get  
 \begin{equation}
 K_jL = 2\pi j - 2L\int dp \rho(p) \tan^{-1}[(K_j -p)/c].
 \end{equation}
 where the set of $K_j$ is a unification of the sets of $\tilde k$ and $k$. Now we can formally introduce the density of holes: 
 \begin{equation}
  \tilde\rho(K) +\rho(K) = \mbox{lim}_{L\rightarrow \infty} \frac{1}{L(K_{j+1}-K_j)},  
 \end{equation}
 to get the final result:
 \begin{equation}
  \frac{1}{2\pi} =\tilde\rho(k) +\rho(k) - \int dp Q(k-p)\rho(p),   
 \end{equation}
 where the kernel $Q(k) = \frac{c}{\pi(k^2 + c^2)}$. 
 The derivation of Eqs.~(\ref{eq:TBA}) follows the same scheme, and mathematical identities outlined in the next section of this appendix were used in the process. 
 
\subsection{Useful Identities}

Some useful identities involving the kernels introduced in Eq.~\eqref{eq:DefsMainText} 
include
\begin{subequations}
\begin{align}
    a_n(\lambda) &= \frac{i}{2\pi} \frac{d}{d\lambda} \ln e_n(\lambda)= \frac{1}{2\pi}\frac{n}{\lambda^2 + (n/2)^2} &&\rightarrow e^{-n|\omega|/2},\\
    A_{nm}(\lambda) & = \frac{i}{2\pi} \frac{d}{d\lambda} {\sum}'_{\vert n - m \vert \leq k \leq n + m - 1} \ln[e_k(\lambda)e_{k+1}(\lambda)]&&\rightarrow 
    \coth|\omega|/2|\Big(e^{-\frac{|n-m||\omega|}{2}}- e^{-\frac{(n+m)|\omega|}{2}}\Big),\\
    s(\lambda) &= \frac{1}{2\pi}\frac{1}{\cosh(\pi\lambda)}&&\rightarrow \frac{1}{2\cosh(\omega/2)},\\
    C_{nm}(\lambda) &= [A^{-1}]_{nm}= \delta_{nm}\delta(\lambda) - s(\lambda)[\delta_{n,m-1}+\delta_{n,m-1}] .
\end{align}
\end{subequations}
Here, the apostrophe in $\sum'_k$ implies that the running index $k$ is incremented in steps of $2$. Fourier transformation is indicated by an arrow.

It is implicitly understood that matrices $A_{nm}$, $C_{nm}$ are defined to be zero for $n,m \leq 0$. We also highlight that the inverse of $A_{nm}$ assumes summation over all integer $n$. If the summation is restricted over even $n$ for $A_{nm}^{(even)}(\omega) = (1-d)^{-1} A_{n/2,m/2}(2\omega)$ where both $n,m$ are even, we use
\begin{equation}
\tilde C_{mn}(\omega)  = [1- d(\omega)]  [\delta_{nm} - s(2\omega)(\delta_{n, m +2} + \delta_{n, m-2})] = [1- d(\omega)] C_{m/2,n/2}(2\omega),
\end{equation}
which has the property 

\begin{subequations}
\begin{equation}
\sum_{n \text{ even}} \tilde C_{m n} A_{nm'} = \delta_{mm'} + \sum_\pm \delta_{m,m'\pm 1} f(\omega), \label{eq:App:Ceven}
\end{equation}
where
\begin{equation}
f(\omega) = \frac{\sinh(\omega/2)}{\sinh(\omega)}.
\end{equation}
\end{subequations}

\subsection{Inverse of kernel of TBA}
\label{app:InverseKernel}

In order to invert Eq.~\eqref{eq:TBA} we need the inverse $\mathbf{C}_{m'n}^{l'j}(\omega)$ of $\mathbf{A}_{nm}^{jl}(\omega)$, Eq.~\eqref{eq:Kernel}, which we rewrite as

\begin{equation}
\mathbf{A}_{nm}^{jl}(\omega) = \left [(\mathbf 1 - \Pi_{\rm odd} \bar \Pi^{(k)}) \underline{A} * (\underline{\bar C} - d \bar \Pi^{(k)}) (\mathbf 1 - \Pi_{\rm odd} \bar \Pi^{(k)}) \right]_{nm}^{jl}(\omega).
\end{equation}
Here, underlined quantities are matrices which act in the space of string centers (roots) $n,m, = 1,2,\dots, \infty$ ($j,l = 1, \dots k$) if they have no (if they have a) bar over the symbol. $\Pi_{\rm odd}$ is the projector on odd string centers $[\Pi_{\rm odd}]_{nm} = \delta_{n,m} (1 - (-1)^n)/2$ and $[\bar \Pi^{(k)}]_{jl} = \delta_{jk} \delta_{lk}$ projects onto the $k$th root. 

We formally use the Ansatz

\begin{equation}
\mathbf{C}_{nm}^{jl}(\omega) = [\mathcal A_{\rm tot} * [\underline{C} (\mathbf 1 - \bar \Pi^{(k)}) + \Pi_{\rm even} \underline{\tilde C} \bar \Pi^{(k)}  \Pi_{\rm even}] ]_{nm}^{jl} \label{eq:TotalInverseKernel}
\end{equation}
for the inverse of the kernel.

To construct the inverse, we begin by multiplying the kernel as follows (note that the kernel contains only even string centers for the $k$th root, so we impose the same structure for the stepwise construction of the inverse)

\begin{align}
[\underline{C} (\mathbf 1 - \bar \Pi^{(k)}) + \Pi_{\rm even} \underline{\tilde C} \bar \Pi^{(k)}  \Pi_{\rm even}] *\mathbf{A}
& = \Pi_{\rm odd} [(\mathbf 1 -  \bar{\Pi}^{(k)})\underline{\bar C} (\mathbf 1 - \bar{\Pi}^{(k)}) ]\Pi_{\rm odd}  \notag\\
& + \Pi_{\rm even} (\underline{\bar C} - d \bar \Pi^{(k)}) \Pi_{\rm even} \notag \\
& - \Pi_{\rm even} (\Pi_+ + \Pi_-) \Pi_{\rm odd} (f*s) \hat e_{k} \hat e_{k-1}^T
\end{align}
Here, $\hat e_j$ is the unit vector $[\hat e_j]_l = \delta_{jl}$ and $[\Pi_\pm]_{m,m'} = \delta_{m, m' \pm 1}$ and we used Eq.~\eqref{eq:App:Ceven}. Note that the inverse of the $k \times k$ matrix $\bar C_{jl}$ is a generalization of the infinite dimensional case Eq.~\eqref{eq:app:A} 
\begin{align}
[\bar C^{-1}]_{jl} & = 2\coth(\omega/2)\underbrace{\frac{\sinh([k + 1 - \text{max}(j,l)]\omega/2)}{\sinh([k + 1]\omega/2)}}_{\stackrel{k\rightarrow \infty}{\longrightarrow} e^{-\text{max}(j,l)\vert\omega\vert/2}}\sinh(\text{min}(j,l)\omega/2). \label{eq:Cinverse}
\end{align}
We use this expression to invert the first to lines. We find
\begin{align}
{ \mathcal A}^{\rm odd}_{jl}(\omega) & = 2\coth(\omega/2){\frac{\sinh\lbrace[k- \text{max}(j,l)]\omega/2\rbrace}{\sinh(k\omega/2)}}\sinh[\text{min}(j,l)\omega/2], & j,l = 1, \dots, k-1,\\
{ \mathcal A}^{\rm even}_{jl}(\omega) & = 2 \coth(\omega/2)\frac{\cosh\lbrace [k+1 - \max {(j,l)}]\omega/2\rbrace}{\cosh[(k+1) \omega/2]}   \sinh[\min {(j,l)}\omega/2], & j,l = 1, \dots, k .\label{eq:Aeven}
\end{align} 
With the help of these expressions we find
\begin{align}
\mathcal A^{\rm tot} &=  [ \underline{\mathcal A^{\rm odd}}^{-1} \Pi_{\rm odd} + \underline{\mathcal A^{\rm even}}^{-1}  \Pi_{\rm even} -\Pi_{\rm even} (\Pi_+ + \Pi_-) \Pi_{\rm odd} (f*s) \hat e_{k} \hat e_{k-1}^T]^{-1} \notag \\
& = \underline{\mathcal A^{\rm odd}} \Pi_{\rm odd} + \underline{\mathcal A^{\rm even}}  \Pi_{\rm even} 
+ \underline{\mathcal A^{\rm even}}  \Pi_{\rm even} (\Pi_+ + \Pi_-) \Pi_{\rm odd} (f*s) \hat e_{k} \hat e_{k-1}^T \underline{\mathcal A^{\rm odd}} . \label{eq:DefAMathcal}
 \end{align}
Note that this is an exact expression (not only the leading order expansion), it can be readily seen that higher order terms in the Taylor series vanish due to $\Pi_{\rm odd} \Pi_{\rm even} = 0$. Here $\hat e_l$ is the basis vector with unit $l$th component.

We will specifically also need 
\begin{subequations}
\begin{align}
\mathcal A^{\rm even}_{j,1}(\omega) & = 2 \cosh(\omega/2) \frac{\cosh[(k+1-j)\omega/2]}{\cosh([k+1] \omega/2)} , \\
\mathcal A^{\rm odd}_{j,1}(\omega) & = 2 \cosh(\omega/2) \frac{\sinh[(k-j)\omega/2]}{\sinh(k \omega/2)} , \\
[f*s* \mathcal A^{\rm even}_{j,k}*\mathcal A^{\rm odd}_{k-1,1}] (\omega) & =  2 \cosh(\omega/2) \frac{\sinh \left(\frac{j \omega }{2}\right)}{2 \sinh \left(\frac{k \omega }{2}\right) \cosh \left(\frac{1}{2} (k+1) \omega \right)}.
\end{align}
\label{eq:A1s}
\end{subequations}

\section{Zero temperature solution at finite magnetic field}
\label{app:ZeroT}

\subsection{Solution of the Wiener-Hopf equation} \label{app:WienerHopf}

Here we remind the reader how to solve Wiener-Hopf equations. The general form of such equation is
\begin{subequations}
\begin{align}
\int_{-\infty}^0
K(x-y)f(y) dy = g(x), ~~ x<0.
\end{align}
\end{subequations}
We may safely assume that $g(x)$ is defined on the entire real axis. As the first step we expand this equation to the entire real axis by introducing function $f^{(+)}(x)$ which vanishes at $x<0$:
\begin{subequations}
\begin{align}
f^{(+)}(x) + \int_{-\infty}^0
K(x-y)f^{(-)}(y) dy = g(x).
\end{align}
\end{subequations}
By Fourier transforming both parts of the equation we obtain
\begin{subequations}
\begin{align}
f^{(+)}(\omega) +K(\omega)f^{(-)}(\omega) = g(\omega), \label{WF2}
\end{align}
\end{subequations}
where now the functions $f^{(\pm)}(\omega)$ are analytical in the lower and upper half plane of $\omega$ respectively. The solution is based on the fact that any integrable function $F(\omega)$ defined on the real axis can be decomposed into a sum of the functions analytic in upper and lower half planes respectively:
\begin{subequations}
    \begin{align}
    F(\omega) &= F^{(+)}(\omega) + F^{(-)}(\omega),\\
    F^{(\pm)}(\omega) &= \pm\frac{i}{2\pi}\int_{-\infty}^{\infty}\frac{d \omega' F(\omega')}{\omega -\omega' \pm i0}.\label{decomp}
    \end{align}
\end{subequations}
Applying this to (\ref{WF2}) we decompose the kernel into the product of functions analytic in opposite half planes: $K(\omega) = G^{(+)}(\omega)G^{(-)}(\omega)$ and rewrite it as 
\begin{subequations}
\begin{align}
f^{(+)}(\omega)/G^{(+)}(\omega) + f^{(-)}(\omega)G^{(-)}(\omega) = g(\omega)/G^{(+)}(\omega),
\end{align}
\end{subequations}
Then according to (\ref{decomp}) the solution of (\ref{WF2}) is 
\begin{subequations}
    \begin{align}
 f^{(+)}(\omega) &= \frac{i G^{(+)}(\omega)}{2\pi}\int \frac{d\omega' g(\omega')}{G^{(+)}(\omega')(\omega -\omega'+i0)}, \\ 
 f^{(-)}(\omega) &= -\frac{i}{2\pi G^{(-)}(\omega)}\int \frac{d\omega' g(\omega')}{G^{(+)}(\omega')(\omega -\omega'-i0)}. \label{WH3}
 \end{align}
\end{subequations}

\subsection{Inversion of the TBA equations}
As highlighted in the main text, at $T = 0$ only  
$\rho_2^{(j)} \neq 0$ ($j <k$) and $\sigma_1 = \rho_2^{(k)} \neq 0$, so it is possible to concentrate on Eq.~\eqref{eq:TBA} keeping only $n = m = 2$. Then the inverse TBA equations can be obtained by using 
\begin{equation}
K_{jl}(\omega) = (\mathbf{A}_{22})^{-1}_{jl}(\omega) = \mathcal A_{jl}^{\rm even}(\omega)/A_{22}(\omega),
\end{equation},
with $\mathcal A_{jl}^{\rm even}$ given in Eq.~\eqref{eq:Aeven}. This leads to eq.~\eqref{eq:ZeroTInverseTBA} of the main text.

\subsection{Rescaling, Wiener-Hopf and scaling limit}
It is convenient to count the excitations relative to the Fermi energy
\begin{subequations}
\begin{align}
\rho^{(+)}_j(\lambda) &= \rho_2^{(j)}(\lambda-B_j),\\
\rho^{(-)}_j(\lambda) &= \tilde \rho_2^{(j)}(\lambda-B_j),
\end{align}
\end{subequations}
so that $\rho^{(\pm )}_j(\lambda) = \int (d\omega) e^{- i \omega \lambda} \rho_j^{(\pm)}(\omega) \propto \theta(\pm \lambda)$ if $\rho^{(+ )}_j(\omega)$ [$\rho^{(- )}_j(\omega)$] are analytical in the upper (lower) half-plane.

With this notation Eq.~\eqref{eq:ZeroTInverseTBA} evaluated at $\lambda - B_j$

\begin{align}
\rho^{(+)}_j(\lambda) + \int_{-\infty}^0 d\lambda' K_{jl}(\lambda - \lambda ' +B_l- B_j) \rho^{(-)}_l(\lambda') & = r_0^{(j)}(\lambda - B_j) + \frac{1}{N} r_1^{(j)} (\lambda +1/g- B_j) \notag\\
& = \frac{2}{1+k} \sin\left ( \frac{\pi l}{2k+2}\right) e^{\frac{\pi (\lambda -B_j)}{1+k}} + \frac{1}{N} r_1^{(j)} (\lambda +1/g- B_j). \label{eq:ScalingWH}
\end{align}
We will be interested in the limit of large $B_j$, and thus for finite $\lambda$ we can consider the scaling limit, in which we keep only the leading residue of the convolution $r_0^{(j)}$ defined by
\begin{subequations}
\begin{align}
[K_{j1} *A_{22}*s](\lambda) & = r_0^{(j)}(\lambda) \leftrightarrow r_0^{(j)} (\omega) = \frac{\cosh[(k + 1 - j) \omega/2]}{\cosh[(k + 1) \omega/2]}, \label{eq:app:r0}\\
[K_{j1} *A_{21}*s](\lambda) & = r_1^{(j)}(\lambda) \leftrightarrow r_1^{(j)} (\omega) = r_0^{(j)}(\omega) \frac{\sinh(q \omega/2)}{\sinh(\omega)} \label{eq:app:r1}.
\end{align}
\end{subequations}
This corresponds to taking the universal limit 
\begin{equation}
E_F = 1/\pi a_0\rightarrow \infty, ~~ g \rightarrow 0, ~~T_K = \mbox{const}.\label{scaling}
\end{equation}
when higher powers of $h/E_F$ are discarded. 

\subsection{Limit of a single non-zero $B_j$}

We first consider the case when only one of the $B_j>0$ is finite, while $B_l \rightarrow \infty$ for $l \neq j$.

\begin{align}
\rho^{(+)}_j(\lambda) + \int_{-\infty}^0 d\lambda' K_{jj}(\lambda - \lambda ') \rho^{(-)}_j(\lambda') & =   e^{\frac{-\pi B_j}{1+k}}r_0(\lambda) + \frac{1}{N} r_1^{(j)} (\lambda +1/g- B_j) \notag\\
\rho^{(+)}_j(\omega) +  K_{jj}(\omega) \rho^{(-)}_j(\omega) & =  \ e^{\frac{-\pi B_j  }{1+k}}r_0(\omega) + \frac{e^{i \omega (B_j - 1/g)}}{N} r_1^{(j)} (\omega)
\end{align}

We next use the Wiener-Hopf method, \ref{app:WienerHopf}, splitting the kernel $K_{jj}(\omega) = G_j^{(+)}(\omega)G_j^{(-)}(\omega)$ into functions which are analytical in upper/lower half-plane.
For example for $k = 2, j = 1$ we find 
\begin{equation}
    G_1^{(-)}(\omega) = \frac{\Gamma(1/2 +i\omega/2\pi)\Gamma(1/2 +3i\omega/2\pi)\exp\{-i(\omega/\pi)\ln[3^{3/2}(\omega -i0)/4\pi e)]\}}{\sqrt 2 \Gamma(1/2)\Gamma(1/2 +i\omega/\pi)}.
\end{equation}

This function satisfies $G(\omega \rightarrow \infty) = 1$. Using this
\begin{equation}
\frac{\rho^{(+)}_j(\omega)}{G_j^{(+)}(\omega)} +  G_j^{(-)}(\omega) \rho^{(-)}_j(\omega)  =  \frac{ e^{\frac{-\pi B_j  }{1+k}}r_0(\omega)}{{G_j^{(+)}(\omega)} } + \frac{\frac{e^{i \omega (B_j - 1/g)}}{N} r_1^{(j)} (\omega)}{{G_j^{(+)}(\omega)} }.
\end{equation}
The solution to this equation is by construction analytical in the lower half-plane
\begin{equation}
G_j^{(-)}(\omega) \rho^{(-)}_j(\omega)  =  \frac{ e^{\frac{-\pi B_j  }{1+k}} r_0^{\rm scaling,(-)}(\omega)}{{G_j^{(+)}(\omega)} } - \frac{i}{2\pi N}\int d \omega' \frac{1}{ \omega - \omega' - i0}\frac{e^{i \omega' (B_j - 1/g)} r_1^{(j)} (\omega')}{{G_j^{(+)}(\omega')} },
\end{equation}
and
\begin{equation}
r_0^{\rm scaling,(-)}(\omega) = \frac{2}{k+1} \sin\left (\frac{\pi j}{2(k+1)}\right) \frac{1}{\frac{\pi}{k+1} + i \omega}.
\end{equation}

The total number of holes $\rho^{
(-)}(\omega = 0) = \int d\lambda \tilde \rho_2^{(j)} (\lambda - B_j)$ determines the leading order bulk
change in magnetization
\begin{equation}
\delta M_j =  M_j ( h_j)-M_j ( 0) = -\rho^{(-)}(\omega \rightarrow 0) =  -A_j e^{\frac{-\pi B_j  }{1+k}} \label{eq:DeltaM}
\end{equation}
for some constant $A_j$.
On the other hand given our definition of the energy $E = h_j M_j$ we see that $M_j \sim -h_j a_0$ and thus $B_j = - (1+k)\ln(h_j a_0)/\pi$ and
\begin{equation}
\frac{1}{g} = -\frac{1+k}{\pi} \ln(a_0 T_K^{(M)}),
\end{equation}
so that $B_j - 1/g = - (1+k)\ln(h_j /T_K^{(M)})/\pi$. We recall the reader that the Kondo temperature entering different physical observables may differ by a universal numerical coefficient, for example in the derivation of the specific heat below, one may have a slightly different definition of the Kondo temperature. 

With this notation we find the impurity magnetization, Eq.~\eqref{eq:Mag} of the main text. 
For $\ln(T_K^{(M)}/h_j)>0$ we close the contour in the upper half plane and consider the residues of

\begin{equation}
r_1^{(j)} (\omega) = \frac{\cosh[(k + 1 - j) \omega/2]}{\cosh[(k + 1) \omega/2]}\frac{\sinh(q \omega/2)}{\sinh(\omega)} = \frac{\cosh[(k + 1 - j) \omega/2]}{\cosh[(k + 1) \omega/2]} \begin{cases} \frac{1}{2 \cosh(\omega/2)}, & q =1, \\
1, & q =2,\end{cases}
\end{equation}

which are at $\omega = i \pi \frac{1}{k+1} (1 + 2n)$, $\omega = i \pi  (1 + 2m)$ (the latter are absent for $j =k$). For even $k$ there are obvious double poles starting at $n =k/2$ and $m= 0$. 
For odd $k$ there are no double poles, but the $m = 0$ pole leads to a power law.

 In summary we find the leading magnetization, 
\begin{align}
M_j\vert_{\rm imp} &\simeq -2 \sec \left(\frac{\pi }{2 k+2}\right) \sin \left(\frac{\pi  j}{2 k+2}\right) \frac{1}{2\pi  G_j^{(-)}(\omega = 0) G_j^{(+)}[\omega = i \pi/(k+1)]} \frac{h_j}{T_K} , \label{eq:app:M}
\end{align}
and non-analytical contributions are of the order $(h_j/T_K)^{k+1} \ln(T_K/h_j)$, exist only for even $k$ and $j <k$, while the non-analytical contributions are $(h_j/T_K)^{k+1}$ without additional contribution at odd $k$.

\subsection{Two non-zero $h_{k} \gg h_l >0$ with $l<k$}

We return to the TBA Equations Eq.~\eqref{eq:TBA}, fix $n = m = 2$ and use $B_k \gg B_j$. For the solution of $\rho_k, \tilde \rho_k$ we use the results from the previous section, while $\tilde \rho_2^{(j)}$ for $j = 1, \dots, k$ are extremely small. We can thus perturbatively solve for $j = 1, \dots k-1$ 

\begin{align}
\delta_{j,1} \left [A_{2,1}(\lambda) + \frac{1}{N} 
[A_{2,q} * s](\lambda + 1/g) \right ]  = \tilde \rho_2^{(j)} (\lambda) + \sum_{l = 1}^k [{A}_{22} * C^{jl}* \rho_{2}^{(l)}](\lambda).
\end{align}
Using the inverse of the $(k-1)\times(k-1)$ matrix $C_{jl}$, cf.~\eqref{eq:Cinverse},
\begin{equation}
\mathcal{\bar A}^{jl} = \frac{2 \coth(\omega/2)}{\sinh(k\omega/2)} \sinh([k- \max(j,l)]\omega/2) \sinh(\min(j,l) \omega/2),
\end{equation}
leading to Eq.~\eqref{eq:barK} with $\bar K_{jl} (\omega)= \bar{\mathcal A}_{jl}(\omega)/A_{22}(\omega) $ and
\begin{subequations}
\begin{align}
\bar r_0^{(j)}(\omega) &= [\bar K_{j1}*A_{22}*s](\omega) = \frac{\sinh[(k-j) \omega/2]}{\sinh(k \omega /2)},\\
\bar r_1^{(j)}(\omega) &=\bar r_0^{(j)}(\omega) \frac{\sinh(q \omega/2)}{\sinh(\omega)},
\end{align}
\end{subequations}
we find $\sum_{j =1}^{k-1} \mathcal{\bar A}^{lj} C_{jk} = - \bar r_0^{(k-j)}$ and thus
\begin{equation}
\rho_2^{(l)}(\lambda) + \bar K_{lj} * \tilde \rho_2^{(j)} (\lambda) = \bar r_0^{(l)} (\lambda) + \frac{1}{N} r_1^{(l)} (\lambda + 1/g) + [\bar r_0^{(k-l)}* \rho_2^{(k)}] (\lambda).
\end{equation}
We now again focus on a single non-infinite $B_l$, as before we introduce particle and hole densities relative to the ``Fermi level''
\begin{subequations}
\begin{align}
\rho^{(+)}_l(\lambda) &= \rho_2^{(l)}(\lambda-B_l),\\
\rho^{(-)}_l(\lambda) &= \tilde \rho_2^{(l)}(\lambda-B_l),
\end{align}
\end{subequations}
and consider the scaling limit, in which we only keep only the leading pole of 
\begin{equation}
\bar r_0^{(l)}(\omega) \simeq \frac{2i}{k} \frac{\sin\left (\frac{(k-l) \pi}{k} \right)}{\omega - i 2\pi/k}.
\end{equation}
We thereby find
\begin{equation}
\rho^{(+)}_l(\omega) + \bar K_{ll}(\omega) \tilde \rho^{(-)}_l (\lambda) = e^{- 2\pi B_l/k}\bar r_0^{(l)} (\omega) \left [1 + e^{2\pi B_k/k} \rho_2^{(k)} (\omega) \right]+ \frac{e^{i \omega(B_l-1/g)}}{N} r_1^{(l)} (\omega),
\end{equation}
where to leading order in $1/N$
\begin{equation}
 e^{2\pi B_k/k}\rho_2^{(k)} (\omega \rightarrow 0) = - A_k e^{- B_k \left [\frac{\pi}{k+1} - \frac{2\pi}{k} \right]} \sim - \left (\frac{h_k}{E_F} \right)^{- \frac{2 + k}{k}} \ll -1.
\end{equation}
Here we used Eq.~\eqref{eq:DeltaM} to relate $B_k$ and $h_k$ and employ the analogous logic to find the relationship
\begin{equation}
B_l = - \frac{k}{2\pi} \ln \left [\frac{h_l}{E_F}\left (\frac{h_k}{E_F} \right)^{\frac{2 + k}{k}}\right] \equiv- \frac{k}{2\pi} \ln \left [\frac{h_l^{\rm eff}}{E_F}\right].
\end{equation}
Using the Wiener-Hopf method and $\bar K_{ll}(\omega) = D^{(+)}_l(\omega)D^{(-)}_l(\omega)$ we find (analogously to Eq.~\eqref{eq:app:M}) Eq.~\eqref{eq:Mag2} of the main text.

\section{Finite temperature solution}
\label{sec:DerivationTBAEquations}

\subsection{General strategy}
In this Appendix we remind the reader the basic facts about the Bethe ansatz thermodynamics. At thermal equilibrium the equations for densities are augmented by equations for their ratios: 
 \[
\epsilon_n = T\ln(\tilde\rho_n/\rho_n)
\]
The quantities $\epsilon_n$ which can be interpreted as excitation energies are determined by the  minimization the free energy
\begin{subequations}
\begin{align}
    F &= E[\{\rho,\tilde\rho\}] -TL\sum_n \int d\lambda\Big[(\tilde\rho_n + \rho_n)\ln(\tilde\rho_n +\rho_n) \nonumber\\
    & -\rho_n\ln\rho_n - \tilde\rho_n\ln\tilde\rho_n\Big],
    \end{align}
\end{subequations}
with respect to variations $\delta\rho_n, \delta\tilde\rho_n$ subject to the constraint imposed by the Bethe ansatz equations for the densities (\ref{eq:TBA}). 

 The most remarkable property of all gapless chiral theories is the relations between the equilibrium densities and $\epsilon$'s, derived in~\cite{FilyovWiegmann1981}. 
 For  the right-moving particles we have:
 \begin{subequations}
     \begin{align}
 \tilde\rho_n &= \frac{1}{2\pi v}\frac{\partial\epsilon_n}{\partial\lambda}[1- N_F(\epsilon_n)], \nonumber\\
 \rho_n &= \frac{1}{2\pi v}\frac{\partial\epsilon_n}{\partial\lambda}N_F(\epsilon_n), \label{relations}
 \end{align}
 \end{subequations}
 where $N_F(\epsilon)$ is the Fermi distribution function and $v$ is the velocity of the excitations.
 
 Substituting (\ref{relations}) into the expression for the entropy we obtain after some algebra 
 \begin{subequations}
 \begin{align}
 S/L = -\frac{T}{2\pi v} \sum_n \int_{N_n(-\infty)}^{N_n(+\infty)}d N\Big[\frac{\ln N}{1-N} +\frac{\ln(1-N)}{N}\Big].
 \end{align}
 \end{subequations}
 This relation maintains that the Sommerfeld coefficient in the specific heat is determined by the asymptotic values of the Fermi distribution functions for the BA strings. These asymptotic values can be extracted without solving the entire equations. Since the Sommerfeld coefficient is related to the chiral CFT central charge through the famous formula by Cardy:
 \begin{subequations}
 \begin{align}
 \gamma = \frac{\pi}{6v} C.
 \end{align}
 \end{subequations}
 this allows one to extract the central charge of the part of the bulk coupled to the impurity.
 
 The distinct feature of the Fermi liquid impurities is that the impurity contributions to the densities $\rho_n$ have the same low energy asymptotic as the corresponding bulk densities. Hence the impurity contribution for the  specific heat at low $T$ is also linear in $T$ like in the bulk Fermi liquid.
 
 \subsection{Free energy minimum}
 To derive the full finite temperature TBA Equations, Eqs.~\eqref{eq:PhiEq} of the main text, we use the inverted kernel $\mathbf{C}_{mn}^{jl}$, derived in \ref{app:InverseKernel}
 as well as the
free energy is $F = E - TS$ where we use Eq.~\eqref{eq:EnergyContinuum} and the entropy is
\begin{equation}
S = N \sum_{n} \sum_{j = 1}^k\int d \lambda  (\rho_n^{(j)} + \tilde \rho_n^{(j)}) \ln(\rho_n^{(j)} + \tilde \rho_n^{(j)}) - \rho_n^{(j)}  \ln(\rho_n^{(j)}) - \tilde \rho_n^{(j)} \ln(\tilde \rho_n^{(j)}). \label{eq:DefEntropy}
\end{equation} 
 The variation of Eqs.~\eqref{eq:TBA} with respect to $\rho_n^{(j)}$ leads to 
\begin{equation}
\frac{\delta \tilde \rho_n^{(j)}(\lambda)}{\delta  \rho_m^{(l)}(\lambda')}  = - {\mathbf A}_{nm}^{jl}(\lambda - \lambda').
\end{equation}

Thus, the variation of the entropy is 
\begin{align}
\frac{\delta S}{N} = \int d\lambda \delta \rho_n^{(j)} [\ln (1 + e^{\epsilon_n^{(j)}/T}) - {\mathbf A}_{nm}^{jl} *\ln (1 + e^{-\epsilon_m^{(l)}/T})]
\end{align} 

and the variation of the free energy $F = E - TS$ leads to
\begin{align}
0 =  \delta_{j,1}\frac{p_{n + 1} + \delta_{n \neq 1} p_{n-1}}{2a_0} - h^{(j)} n + T[\ln (1 + e^{\epsilon_n^{(j)}/T}) - {\mathbf A}_{nm}^{jl} *\ln (1 + e^{-\epsilon_m^{(l)}/T})] \label{eq:FreeEnergyMin}
\end{align}
where  we introduced $\rho_{n}^{(j)}/\tilde \rho_n^{(j)} = e^{ - \epsilon_n^{(j)}(\lambda)/T}$.

To find the relationship to magnetic fields, we first multiply this equation by $1/n$ and take the $n \rightarrow \infty$ limit. The contribution of $p_{n \pm 1}$ vanishes and we see that a consistent solution, in which all terms $\ln(1 + e^{- \epsilon_n^{(j)}/T})$ vanish at large $n$, is that
\begin{equation}
\lim_{n\rightarrow \infty}\frac{\epsilon_n^{(j)}}{n} = h^{(j)} >0.
\end{equation}

\subsection{Derivation of universal TBA}

We next invert Eq.~\eqref{eq:FreeEnergyMin} using the results for $\mathbf{C}_{mn}^{jl}$ derived in~\ref{app:InverseKernel} and use that 
\begin{align}
\sum_n C_{mn} * n &= \int d\lambda C_{mn}(\lambda - \lambda') \delta(\lambda') n = m - s(\omega = 0)[m+1+\delta_{m>1}(m-1)] = 0,\\
\sum_n' \tilde C_{mn} * n &= \int d\lambda \tilde C_{mn}(\lambda - \lambda') \delta(\lambda') n = [1-d(\omega = 0)] [m - s(\omega = 0)[m+2+\delta_{m>2}(m-2)]] =0
\end{align}
and 
\begin{align}
p_{n + 1}(\omega) + \delta_{n \neq 1} p_{n - 1}(\omega) 
&= 2\pi [A_{n,1} * \theta - \delta_{n,1} \theta] (\omega)
\end{align}
to get the partially inverted equation of $\epsilon_n$

\begin{align}
0 =  \frac{\pi \delta_{j,1}}{a_0}\underbrace{[\delta_{m,1} \theta - C_{m,1} * \theta]}_{= s*\theta\delta_{m,2}} + T[(C_{mn} \delta_{j \neq k}+\tilde C_{mn} \delta_{j k})*\ln (1 + e^{\epsilon_n^{(j)}/T}) - 
[\tilde{\mathcal A}_{\rm tot}^{-1}]_{jl}*\ln (1 + e^{-\epsilon_m^{(l)}/T})]. \label{eq:FreeEnergyMinPartialInv}
\end{align}

Next we use Eq.~\eqref{eq:DefAMathcal} to invert the equation fully and exploit $\mathcal A_{l,1}*s*\theta\delta_{m,2}  = r_0^{(l)}*\theta \delta_{m,2}$ with the Fourier transform of $r_0^{(l)}(\omega)$, Eq.~\eqref{eq:app:r0} is
\begin{equation}
r_0^{(l)}(\lambda)=\frac{2}{1+k} \frac{\cosh\left(\frac{\pi \lambda}{k+1}\right)\sin\left(\frac{\pi l}{2k+2}\right)}{\cosh\left(\frac{2\pi \lambda}{k+1}\right)-\cos\left(\frac{\pi l}{k+1} \right)} \stackrel{\vert \lambda \vert \gg k+1}{\simeq} \frac{2}{1+k} \sin\left ( \frac{\pi l}{2k+2}\right) e^{- \frac{\pi \vert \lambda \vert}{1+k}}. \label{eq:r0inverse}
\end{equation}
Here, the symbol $\simeq$ denotes the scaling limit, which we use to study the fully inverted equation
\begin{equation}
0 = r_0^{(l)} * \theta \frac{\delta_{m,2} \pi}{a_0} + T \mathcal A^{\rm tot}_{lj}*[(C_{mn} \delta_{j \neq k}+\tilde C_{mn} \delta_{j k})*\ln (1 + e^{\epsilon_n^{(j)}/T}) - 
[\tilde A_{\rm tot}^{-1}]_{jl'}*\ln (1 + e^{-\epsilon_m^{(l')}/T})].
\end{equation}
at $\lambda \rightarrow \lambda + \frac{1+k}{\pi}\ln(a_0 T/2) \ll -1$. We use 
\begin{align}
r_0^{(l)} * \theta [\lambda + \frac{1+k}{\pi}\ln( a_0 T/2)] & = \int^{\lambda + \frac{1+k}{\pi}\ln( a_0 T/2)}_{-\infty} d\lambda' r_0^{(l)}(\lambda') 
\simeq \frac{ a_0 T}{\pi}\sin\left ( \frac{\pi l}{2k+2}\right) e^{\frac{\pi  \lambda }{1+k}}
\end{align}
and define $\epsilon_n^{(j)}[\lambda + \frac{1+k}{\pi}\ln( a_0 T/2)]/T \equiv \phi_n^{(j)}(\lambda)$ to get

\begin{align}
- \sin\left ( \frac{\pi l}{2k+2}\right) e^{\frac{\pi  \lambda }{1+k}}\delta_{m,2} &=  \mathcal A^{\rm tot}_{lj}*[(C_{mn} \delta_{j \neq k}+\tilde C_{mn} \delta_{j k})*\ln (1 + e^{\phi_n^{(j)}}) - 
[\tilde A_{\rm tot}^{-1}]_{jl'}*\ln (1 + e^{-\phi_m^{(l')}})]\notag\\
& = \phi_m^{(l)} + [\mathcal A^{\rm tot}_{lj}*(C_{mn} \delta_{j \neq k}+\tilde C_{mn} \delta_{j k})- \delta_{lj} \delta_{mn}]*\ln (1 + e^{\phi_n^{(j)}}).
\end{align}
This concludes the derivation of Eq.~\eqref{eq:PhiEq} of the main text.

\subsection{Free energy}
\label{app:Freeenergy}

Next, we rewrite the free energy using the inversion of Eq.~\eqref{eq:TBA} by means of $\mathbf C_{nm}^{jl}$ derived in~\ref{app:InverseKernel} 
\begin{align}
\frac{F}{N} &=  - \sum_{j = 1}^k\sum_n \int d \lambda   \left [\frac{p_{n+1}(\lambda) + (1-\delta_{n1})p_{n-1}(\lambda)}{2a_0} - h_j n/N - T \ln\left(1 + \tilde \rho_n^{(j)}/\rho_n^{(j)}\right) \right ]{\rho_n^{(j)}(\lambda)}
\notag \\
&- T\sum_{j = 1}^k\sum_n \int d \lambda    \ln\left (1 + \rho_n^{(j)}/\tilde \rho_n^{(j)}\right) \tilde \rho_n^{(j)}(\lambda) \notag\\
&= - T \int d\lambda \ln( 1 +  e^{\epsilon_m^{(l)}(\lambda)/T} ) \left [ [{\mathbf C}_{mn}^{l1} *A_{n,1}](\lambda) + \frac{1}{N} \begin{cases} [{\mathbf C}_{mn}^{l1} * A_{n,1} * s](\lambda + 1/g) 
\\  [{\mathbf C}_{mn}^{l1} * A_{n,1} ](\lambda + 1/g) 
\end{cases} \right ] . \label{eq:app:F}
\end{align}

{The leading temperature dependence at smallest temperatures stems from the logarithm multiplying the square bracket, which in turn is $\rho_{m}^{(l)} +{\mathbf C}_{mn}^{lj} \tilde \rho_{n}^{(j)}$, see Eq.~\eqref{eq:TBA}. As a simple way to obtain the bulk contribution, we thus concentrate on the contribution of $\rho_{n}^{(j)} \simeq \delta_{n,2} r_0^{(j)}$, see Eqs.~\eqref{eq:ZeroTInverseTBA} and \eqref{eq:app:r0}. Note that both the functional dependence of bulk contribution and the impurity is identical.}

In the scaling limit, we shift the integration variable to $\lambda \rightarrow \lambda +  (1+k)\ln ( T/E_F)/\pi$, use the notation $\phi_n^{(j)}$ introduced above and use $(1+k)\ln ( T/E_F)/\pi + 1/g \equiv (1+k)\ln(T/T_K)/\pi$ with 
\begin{equation}
\frac{1}{g} = -\frac{1+k}{\pi} \ln(T_K/E_F) \Leftrightarrow T_K =E_F e^{-\pi/g(1+k)}.
\end{equation}
{This leads to Eqs.~\eqref{eq:FimpFiniteT},~\eqref{eq:GFMaintext} of the main text. The leading non-Fermi-liquid corrections for $q = 1$ stem from $n = 1$ in Eq.~\eqref{eq:app:F}}
where we define 
\begin{align}
	g_j^{(1)}(\omega)& ={\mathbf C}_{1n}^{l1} * A_{n,1} * s = \mathcal A_{j1}^{\rm odd}*s =\frac{\sinh[(k-j)\omega/2]}{\sinh(k \omega/2)}.
\end{align}
 
\subsection{Zero temperature limiting equations and low-temperature entropy.}
\label{app:ZeroTDeg}
We now extract the behavior of $\phi_1^{(j)}$ at $\lambda \rightarrow \infty$, which governs the limit of zero temperature. To this end, we first study $\phi_2^{(j)}$ as the largest term, the source term $-c_j e^{\pi \lambda/(1+k)}$ on the left, needs to be balanced by the $\phi_2^{(j)} (\lambda) \stackrel{\lambda \rightarrow \infty}{\simeq} - c_j e^{\pi \lambda/(1+k)}$. 

Next we consider the k-1 equations for $m = 1$, where we can drop terms $\mathcal {O}e^{- \pi \lambda/(1+k)}$
\begin{align}
0 &= \phi_1^{(j)} + s *\ln [(1 + e^{\phi_{1}^{(j+1)}})(1 + e^{\phi_{1}^{(j-1)}})] \notag\\
& \rightarrow \phi_{1,\infty}^{(j)} + \frac{1}{2}\ln [(1 + e^{\phi_{1,\infty}^{(j+1)}})(1 + e^{\phi_{1,\infty}^{(j-1)}})],
\end{align}
where the second line assumes constant $\phi_{1}^{(j)}$ in the $\lambda \rightarrow \infty$ limit.
The solution thereof parallels the $k-$channel SU(2) Kondo problem~\cite{Tsvelick1985}
\begin{align}
(1 + e^{- \phi_{1,\infty}^{(j)}}) & = \frac{\sin^2\left(\frac{\pi (j + 1)}{k + 2}\right)}{\sin^2\left(\frac{\pi}{k + 2}\right)}
\end{align}
which is even under $j \rightarrow k-j$.

For the leading temperature corrections, we highlight that they are determined by the leading residue of $F_j^{(2)}$, which is $\omega = i \pi/(k+1)$ with residue and that $F_j^{(2)}(\omega)/F_j^{(1)}(\omega) = 2 \cosh(\omega/2)$. Therefore, 
$\text{Res}[F_j^{(2)}(\omega)] /\text{Res}[F_j^{(1)}(\omega)] = 2 \cos \left(\frac{\pi }{2 k+2}\right)$.

\bibliography{SymplecticBethe}

\end{document}